\pdfoutput=1
\documentclass[usenatbib]{mn2e}
\usepackage{hyperref}
\usepackage{amsfonts,amsmath,amssymb}
\usepackage{txfonts}
\usepackage{graphicx}
\usepackage{bm}
\usepackage{ctable}
\usepackage{url}
\usepackage{breakurl}
\usepackage{fixltx2e} 
\usepackage[]{cleveref}
\UseRawInputEncoding
\crefname{section}{\S\!}{\S\S\!}
\crefname{appendix}{App.}{Apps.}
\crefname{equation}{Eq.}{Eqs.}
\Crefname{equation}{Equation}{Equations}
\crefname{figure}{Fig.}{Figs.}
\crefname{table}{Tab.}{Tabs.}
\Crefname{figure}{Figure}{Figures}
\usepackage{mathtools}
\usepackage{cuted}
\newcommand{\revchng}[1]{{#1}}
\newcommand{\acknowledgements}[1]{\begin{small}\section*{Acknowledgments}\end{small}{\noindent #1}\vspace{10pt}}
\newcommand\altaffilmark[1]{$^{#1}$}
\newcommand\altaffiltext[1]{$^{#1}$}
\newcommand\wsj{w_{s,j}}
\newcommand\ws{w_{s}}
\newcommand\ad{\epsilon_{\rm grain}}
\newcommand\ab[1]{\epsilon_{{\rm gr},{#1}}}
\newcommand\abc[1]{\epsilon_{{\rm gr},{#1}}}
\newcommand\amin{\epsilon_{\rm grain}^{\rm min}}
\newcommand\amax{\epsilon_{\rm grain}^{\rm max}}
\newcommand\adj{\epsilon_{{\rm grain},j}}
\newcommand\tsj{t_{s,j}}
\newcommand\ddt[1]{\frac{\partial #1}{\partial t}}
\newcommand\aext{{\rm a}^{\rm ext}}
\newcommand\baext{{\bf a}^{\rm ext}}

\newcommand\hs{HS18}
\newcommand\msh{MSH19}

\newcommand\tfigh{4.5cm}
\title[The RDI with a spectrum of grains]{The Acoustic Resonant Drag Instability with a Spectrum of Grain Sizes\vspace{-0.5cm}}

\author[Squire et al.]{
\parbox[t]{\textwidth}{ 
	Jonathan Squire\altaffilmark{1}, Stefania Moroianu\altaffilmark{2}, \&\ Philip~F.~Hopkins\altaffilmark{3}
} 
\vspace*{6pt} \\
\altaffiltext{1}{Physics Department, University of Otago, Dunedin 9010, New Zealand} \\
\altaffiltext{2}{Department of Applied Physics, Stanford University, Stanford, CA 94305, USA}\\
\altaffiltext{3}{TAPIR, Mailcode 350-17, California Institute of Technology, Pasadena, CA 91125, USA\vspace{-0.3cm}}
}

\date{Submitted to MNRAS, 2021\vspace{-0.6cm}}
\begin{document}
\maketitle

\begin{abstract}
We study the linear growth and nonlinear saturation of the ``acoustic Resonant Drag Instability'' (RDI) when the dust grains, which drive the instability, have a wide, continuous spectrum of different  sizes. This physics is generally applicable to dusty winds driven by radiation pressure, such as occurs around red-giant stars, star-forming regions, or active galactic nuclei. Depending on the physical size of the grains compared to the wavelength of the radiation field that drives the wind, two qualitatively different  regimes emerge. In the case of  grains that are larger than the radiation's wavelength -- termed the \emph{constant-drift} regime -- the grain's equilibrium drift velocity through the gas is approximately independent of grain size, leading to strong correlations between differently sized grains that persist well into the saturated nonlinear turbulence.  For grains that are smaller than the radiation's wavelength  -- termed the \emph{non-constant-drift} regime -- the linear instability grows more slowly than the single-grain-size RDI and only the larger grains exhibit RDI-like behavior in the saturated state. A detailed study of grain clumping and grain-grain collisions shows that outflows in the constant-drift regime may be effective sites for grain growth through collisions, with large collision rates but low collision velocities.    \end{abstract}

\begin{keywords}
instabilities -- turbulence -- ISM: kinematics and dynamics -- galaxies: formation -- stars: winds, outflows -- dust, extinction
\vspace{-1.0cm}
\end{keywords}

\vspace{-1.1cm}

\section{Introduction}


Cosmic dust is ubiquitous across the universe and vital to a wide range of astrophysical processes. By mass, it makes
up around $\sim\!1\%$ of the  interstellar medium (ISM) of galaxies, but its strong coupling to radiation
fields implies it can nonetheless strongly influence gas dynamics and cooling in many situations   \citep{Draine2010}. 
More generally, because around half of the metal content of our galaxy is locked up in dust, it plays 
crucial roles in any process that requires metals or solids {\citep{Whittet1992,Draine2003}}. Notably, dust is almost certainly the key ingredient 
for planet formation and life, supplying the necessary reservoir of solids that provide the seeds to make 
planetesimals in protostellar disks {\citep{Chiang2010}}.

This paper deals with the physics of dust moving through gas, with the interaction between the species mediated by drag forces. Such conditions occur, for example, in dust-radiation-pressure driven winds, where an outflow of dusty gas
 is driven by an anisotropic radiation field that couples strongly to the dust. Such outflows are thought
 to be important in  the evolution of asymptotic giant-branch (AGB) stars (which also produce large quantities 
 of dust; e.g., \citealp{Habing1996,Norris2012}), in feedback processes that help regulate star  formation
 and/or active-galactic nuclei \citep{Scoville2003,Thompson2005,Ishibashi2015}, around supernovae \citep[e.g.,][]{Micelotta2018},
 and in the bulk ISM \citep{Weingartner2001a} and circum-galactic medium (CGM; \citealp{Menard2010}).
As shown by \cite{Squire2018}, this situation -- specifically, when the radiation pressure on the dust is stronger than that on the gas, such that
 the gas outflow is driven indirectly through the drag force from the dust -- is unstable to the ``{Resonant Drag Instability}'' (RDI): small perturbations in the gas or dust will grow in time until they become large, driving turbulence  in the gas
 and strong dust clumping.  \cite{Hopkins2018} (hereafter\defcitealias{Hopkins2018}{\hs}\hs) studied the linear features and growth rates of the RDI
 for the case of nearly neutral grains and neutral gas in outflows, while \cite{Hopkins2018a} generalized these
 results to charged grains in magnetized gas. 
 These results were then extended to the nonlinear regime by  \cite{Moseley2019} (hereafter\defcitealias{Moseley2019}{\msh}\msh) and  \citealt{Seligman2019,Hopkins2020} (in the magnetized regime), who
 studied the turbulence induced by the RDI, constructing simple estimates for 
 its saturation amplitude and other properties.

 However, each of these studies  has allowed for the dust grains to have only 
 a single size.
{ As shown by \cite{Krapp2019,Paardekooper2020,Zhu2021} for} the streaming instability in protoplanetary disks (which is
 part of the RDI family; \citealp{Squire2018a}), a range of grain sizes can 
 strongly influence the instabilities in some regimes.   Thus, for more realistic application to astrophysical fluids and 
 outflows, where the
 range in grain sizes easily spans two orders of magnitude or more  \citep{Draine2010}, 
 we must relax the single-grain-size assumption and better understand the influence 
 of a spectrum of grain sizes on the growth rate and saturation of the RDI. This is the first purpose of the present paper.
 We study both the linear growth rates and nonlinear saturation of the ``acoustic RDI'' (\hs) that applies to uncharged 
 grains and neutral gas, and involves the driving of compressive shocks and sound waves 
 by the drifting dust. We find that depending on whether grains are smaller or larger
 than the wavelength of the accelerating radiation field, the presence of a spectrum of grains either has 
 little effect on the acoustic RDI   or reduces the growth rate and saturation of smaller-scale motions.
 In both cases,  key features of the linear-instability structure persist well into the highly turbulent saturated state.
 
 The second purpose of this paper is to better understand  dust clumping and collisions in 
 RDI turbulence (i.e., the saturated state of the {acoustic RDI}). This is important because outflows are highly 
 dynamic and  often thought to be key sites for grain condensation, coagulation, and fragmentation, and the
  latter two of these processes are strongly influenced by turbulence. While well 
 developed theories exist to describe the how standard gas turbulent motions influence dust collisions 
 \cite[e.g.,][]{Ormel2007,Pan2013,Pumir2016}, the structure of the turbulence and clumping driven by the 
acoustic RDI is quite different to standard turbulence in many ways, because the instability operates 
across all scales of the system simultaneously (\msh). In this context, it is important to consider the RDI with a
wide spectrum of grain sizes (as opposed to the single-grain-size RDI) because the nature of the instability suggests that there
could exist interesting
correlations between differently sized grains in some regimes, and grain clumping and collision  statistics 
 depend strongly on grain sizes \citep{Pan2014a,Blum2018,Mattsson2019}.
 With this in mind, our nonlinear study is designed to compare RDI saturation to forced turbulence with passively advected dust. We  do this by designing ``equivalent'' forced turbulence simulations  (i.e., simulations with parameters
 chosen to match the saturated RDI as closely as possible), allowing an explicit comparison of  the statistics of dust in 
 RDI turbulence with those of passive dust in forced turbulence.
 Given the rather detailed nature of these comparisons, we focus on just two RDI case
 studies at the high numerical resolution, but with parameters that can be applicable 
 to a range of astrophysical scenarios. Depending on the regime, we find that the RDI involves
 a  significantly faster rate of  lower-velocity collisions than forced turbulence, particularly  between 
 grains of different sizes. It also exhibits far stronger clumping of the smallest grains, even when
 simple estimates suggest these small grains should be very well coupled to the gas.

The paper is split in two: first the main exposition; second an extended appendix that
studies analytically the linear behavior of the acoustic RDI with a spectrum of grain sizes. This split was chosen because 
the linear calculations, in which we derive simple analytic expressions for the RDI growth rate 
in most of the important regimes,  are necessarily rather technical. They do provide useful 
understanding of the nonlinear behavior, however, so we will refer to those results throughout. 
The main paper starts in  \cref{sec: rdi description} with a detailed description 
of the problem, model, and numerical setup, particularly focusing  on 
important differences that arise in the dust (quasi-)equilibrium depending on the wavelength 
of the radiation compared to the dust size (\cref{sub: acceleration} and \cref{tab: regime}). The numerical results are presented in \cref{sec: results}, 
starting with a discussion of the broad morphological features of the turbulence and how this differs between
regimes and driving (\cref{sub: results: time evolution}), then followed by a more detailed analysis of 
the grain clumping and collisions (\cref{sub: results: statistics,sub: results: collisions}).

\section{Numerical model and physical setup}\label{sec: rdi description}

\subsection{Dust and gas model}\label{sub: basic model}

We model gas dynamics using the standard neutral fluid equations, 
ignoring magnetohydrodynamic effects and charged grains for simplicity in this study, although such effects are important
for many astrophysical regimes (\hs). Dust is modelled numerically
by treating it as a population of individual particles (the super-particle approach), which interact with the
gas through drag forces that depend on the grain size. 
We use $f_{d}(\ad;\bm{x},\bm{v},t)$ to denote the phase-space density of grains with radius $\ad$, velocity $\bm{v}$, at position 
$\bm{x}$, such that the equations of motion are
\begin{gather}
\ddt{\rho_{g}} + \nabla\cdot(\rho_{g}\bm{u}_{g}) = 0,\label{eq: dt cont}\\
\ddt{\bm{u}_{g}} + \bm{u}_{g}\cdot\nabla\bm{u}_{g} = -c_{s}^{2}\frac{\nabla\rho_{g} }{\rho_{g}}+ \frac{1}{\rho_{g}}\int d\ad d\bm{v}\, f_{d}(\bm{x},\bm{v},\ad) \frac{\bm{v}-\bm{u}_{g}}{t_{s}(\ad,\bm{v})},\label{eq: dt mom}\\
\ddt{f_{d}} + \bm{v}\cdot\nabla f_{d} + \frac{\partial }{\partial\bm{v}} \cdot \left[ \left( -\frac{\bm{v}-\bm{u}_{g}(x)}{t_{s}(\ad,\bm{v})}+ \baext (\ad)\right) f_{d}\right]=0.\label{eq: f dt}
\end{gather}
Given the super-particle approach, the final equation is equivalent to 
modeling individual grains of size $\adj$, with  velocity $\bm{v}_{j}$ and position $\bm{x}_{j}$, with 
\begin{equation}
\ddt{\bm{x}_{j}}=\bm{v}_{j},\quad\ddt{\bm{v}_{j}} = -\frac{\bm{v}_{j} - \bm{u}_{g}(\bm{x}_{j})}{t_{s}(\adj,\bm{v}_{j})} + \baext (\adj)\label{eq: dt v indiv}
\end{equation}
as they move through the gas velocity field, then constructing $f_{d}$ by counting dust particles at each gas position.
Here $\rho_{g}$ is the gas density, $\bm{u}_{g}$ is the gas velocity, $c_{s}$ is the sound speed (the equation of state is taken as isothermal), $t_{s}(\ad,\bm{v})$ is the stopping time (see \cref{eq: stopping time epstein} below), and $\baext$ is an external force from radiation pressure on the grains, which we will take (arbitrarily) to be in the $\hat{\bm{z}}$ direction $\baext=\aext\hat{\bm{z}}$ (see \cref{sub: acceleration}). The 
final term in \cref{eq: dt mom} is the dust ``backreaction'' force -- it the force on the gas from the dust -- which 
is neglected in most studies of dust dynamics. We also use $\langle\cdot\rangle$ to denote the volume average, and 
the subscripts $\perp$ and $\|$  to denote velocities in the directions perpendicular and parallel to the mean drift, respectively (e.g., $\langle u_{g,\|}^{2}\rangle=\langle u_{g,z}^{2}\rangle$, $\langle u_{g,\perp}^{2}\rangle=\langle u_{g,x}^{2}+u_{g,y}^{2}\rangle$). A $\delta$ indicates that the mean is subtracted from a quantity before averaging (e.g., $ \delta v_{d,\|}=v_{d,\|}-\langle v_{d,\|}\rangle$). All quantities will be measured in the frame where the gas is stationary $\langle \bm{u}_{g}\rangle=0$. 

To understand dust dynamics, analyze simulations, and compute linear growth rates, it is  helpful to take the zeroth and first velocity moments
of $f_{d}$, defining 
\begin{equation}
\rho_{d}=\int_{\abc{l}}^{\abc{h}} d\ad d\bm{v}\, f_{d},\qquad \bm{v}_{d} = \frac{1}{\rho_{d}}\int_{\abc{l}}^{\abc{h}} d\ad d\bm{v}\,\bm{v} f_{d},\label{eq: vd rhod defs}
\end{equation}
where the integration limits $\abc{l}$ and $\abc{h}$ can be taken over just a subset of grains (if specified as such; see \cref{sub: dust binning}), or the full distribution (if unspecified). 
We use $\mu_{0}$ to denote the 
total average dust-to-gas-mass ratio $\mu_{0}=\langle \rho_{d}\rangle/\langle \rho_{g}\rangle$,  
and 
\begin{equation}
w_{s}(\ad) = \frac{\langle |\bm{v}_{d}(\ad)-\bm{u}_{g}|\rangle}{c_{s}},\label{eq: ws def}
\end{equation}
to denote the mean drift between dust of size $\ad$ and the gas (here $\bm{v}_{d}(\ad)$ is computed from $\abc{l}=\ad$,
$\abc{h}=\ad+d\ad$ in \cref{eq: vd rhod defs}).

Throughout, we assume Epstein drag, using the simple approximation of  \citep{Draine1979a},
\begin{equation}
t_{s}(\ad,\bm{v}) = \sqrt{\frac{\pi}{8}} \frac{\bar{\rho}_{d,{\rm int}}\ad}{\rho_{g}c_{s}}\left(1+\frac{9\pi}{128}\frac{|\bm{v}-\bm{u}_{g}|^{2}}{c_{s}^{2}} \right)^{-1/2},\label{eq: stopping time epstein}
\end{equation}
where $\bar{\rho}_{d,{\rm int}}$ is the internal grain density. Epstein drag is generally appropriate for astrophysical conditions
in which MHD and charging effects can be neglected as done here (generally, for cooler, denser gas; see, e.g., \hs).

\subsubsection{Grain mass distribution}
 We will assume a simple power law distribution of grain sizes between $\ad=\amin$ and $\ad=\amax$. In 
 order to reduce the number of 
free parameters, we use the standard MRN distribution \citep{Mathis1977}, which postulates that the 
 mass of grains $d\mu$ within logarithmic range of sizes $d\ln\ad$ is 
$d\mu/d\ln\ad\propto \ad^{0.5}$, such that most of the mass is in the largest grains, along with the total dust-to-gas-mass ratio $\mu_{0}\approx0.01$. The original distribution postulates that $\amin\approx5{\rm nm}$ and 
$\amax \approx 0.25 {\rm \mu m}$, although subsequent works have suggested that  
this underestimates significantly the population of small grains and misses a population of larger grains, even in the diffuse ISM \citep{Weingartner2001,Zubko2004,Draine2009}. Less is known about the grain distribution in more dynamic environments (e.g., around AGB stars or the AGN dusty torus{, see, e.g., \citealp{Hoefner2018,Murray2005}}), where our simulations can apply by virtue of their dimensionless nature (see \cref{{sub: mapping to astro}} below). Cursory lower-resolution tests of different 
grain-mass distributions (e.g., small-grain dominated, $d\mu/d\ln\ad\propto \ad^{-0.5}$) have not revealed significant differences, so we shall not explore this in detail. {It is, however, worth noting that for the gas of the protoplanetary-disk streaming instability, the grain 
distribution can affect important details of the linear instability \citep{Paardekooper2020,McNally2021,Zhu2021}, so this issue may be worth revisiting in more detail in future work.}

\begin{table*}
\centering
 \begin{tabular}{|| m{0.48\textwidth} m{0.48\textwidth} ||} 
 \hline
 \emph{Constant drift regime} & \emph{Non-constant drift regime}\\ 
 \hline 
 $ \lambda_{\rm rad} < \ad$, $Q_{\rm abs}\sim 1$ (radiative force on the dust) & $ \lambda_{\rm rad} > \ad $, $Q_{\rm abs}\sim \ad/\lambda_{\rm rad}$  (force on the dust) \emph{or} acceleration of the gas \\ 
  $\aext \propto \ad^{-1}$ & $\aext \sim {\rm const.}$\\ 
   $\ws\sim {\rm const.}$, $t_{s}\propto \ad$ & $\ws\propto \ad^{1/2}$, $t_{s}\propto \ad^{1/2}$ (supersonic drift);  $\ws\propto \ad$, $t_{s}\propto \ad$ (subsonic drift)\\
   Linear instability similar to single-grain-size case & Range of resonant angles changes character of linear instability\\
   Strong  correlations between grains of different sizes in saturated state& Grain correlations broadly similar to externally forced turbulence\\
 \hline
 \end{tabular}
 \caption{Summary of some basic   properties of the grains and the RDI in  the constant- and non-constant-drift regimes.}\label{tab: regime}
\end{table*}

\subsection{Grain acceleration regimes}\label{sub: acceleration}

The acoustic RDI  requires a net drift between grains and the gas as their energy source.  As discussed extensively 
in \hs\ and \citet{Hopkins2018a}, 
such a net drift is expected to occur generically in the presence of 
 radiation fields, which couple to the grains more strongly than to the gas under most conditions (e.g., \citealt{Weingartner2001a,Murray2005}), sourcing the $\baext$ term in \cref{eq: f dt}.
In this radiatively driven situation, two acceleration regimes naturally emerge, applying  respectively to grains smaller or larger than the 
wavelength of the accelerating radiation $\lambda_{\rm rad}$. The different scaling of $\ws(\ad)$
has important implications for the development of instabilities. Some basic properties of the different regimes are summarized in \cref{tab: regime} for quick reference.

A coherent radiation field of energy density $e_{\rm rad}$ causes a grain acceleration $\aext \sim Q_{\rm abs} e_{\rm rad} \ad^{2}/m_{d} c^{2}$, where $m_{d}=4/3\, \pi \bar{\rho}_{d}\ad^{3}$ is the grain's mass. The factor $Q_{\rm abs}$ is the absorption efficiency; $Q_{\rm abs}\sim 1$ if $\ad\gg \lambda_{\rm rad}$, while $Q_{\rm abs}\sim \ad/\lambda_{\rm rad}$ if $\ad\ll \lambda_{\rm rad}$ \citep{Weingartner2001a}. Thus, 
large grains in shorter wavelength radiation fields feel an acceleration $\aext \propto \ad^{-1}$, while small 
grains in longer wavelength radiation fields feel an acceleration that is independent of $\ad$.  In such a driven situation, the (quasi-)equilibrium  occurs when the acceleration and drag forces on grains balance,\footnote{Note
that this is not actually a true equilibrium because it occurs with the 
system as a whole (gas and dust) accelerating linearly at a constant rate, unless $\aext$ can be balanced
by another external force on both gas and dust (e.g., gravity). However, this subtlety does not make 
a difference to the arguments here or to our simulations. The issue is discussed in detail in  \hs. } \emph{viz.,} when \begin{equation}
\frac{\ws(\ad,v_{d})}{t_{s}(\ad,v_{d})}=\aext(\ad).\label{eq: ws to solve}
\end{equation}
 Because $t_{s}(\ad,v_{d})\propto \ad$,
this implies that $\ws$ is independent of grain size if  $\ad\gg \lambda_{\rm rad}$, which we term the \emph{constant drift} regime;
the opposite regime, where $\ws(\ad)$ is a function of grain size (for $\ad\ll \lambda_{\rm rad}$) is termed the \emph{non-constant drift} regime. In the latter case, we see from \cref{eq: stopping time epstein,eq: ws to solve} that 
for subsonic drift $\ws\ll1$, $t_{s}$ in the saturated state is independent of $\ws$ so that $t_{s}\propto\ad$ and $\ws\propto\ad$, while 
for supersonic drift $\ws\gg1$ (the case of more relevance to this article), the decrease in $t_{s}$ with $\ws$ implies that $t_{s}\propto \ad^{1/2}$ and $\ws\propto \ad^{1/2}$.
Of course, in many physical situations, the distribution of grain sizes could fall around $\lambda_{\rm rad}$ (i.e., $\amin<\lambda_{\rm rad}<\amax$), in which case the larger grains will lie in the 
constant-drift regime and the smaller grains in the non-constant-drift regime. However, given that
our goal here is to explore the basic physics of the multi-grain RDI, we will not consider such situations in detail in this work.

A net drift of grains can also be set up through a force on the gas (and not the dust), due to radiation pressure absorbed by gas (e.g., line pressure)  or gravity (e.g., in a stratified medium). In this case, the situation is identical to the non-constant-drift regime. {This physics is applicable, albeit with a 
different linear instability, to the  polydisperse streaming
instability in protoplanetary disks \citep{Krapp2019}. }

{Note that \cite{Hopkins2021} also explore the RDI with a spectrum of grain sizes, explicitly
including vertical stratification and more complex radiation-MHD effects. For some cases, they also consider simulation variants in the $\ad\gg \lambda_{\rm rad}$  and $\ad\ll \lambda_{\rm rad}$
 regimes. Rather than using the labels ``constant-drift'' and ``non-constant-drift,'' they label the $\ad\ll \lambda_{\rm rad}$ (non-constant-drift) regime simulations with ``\textbf{-Q},'' to signify that $Q_{\rm abs}$ depends on $\ad$ in this regime. 
 This different nomenclature is used their because  most of their simulations use
 explicit radiation-MHD effects, which leads to more complex relations between $\ws$ and $\ad$ that depend on the dynamics of
 the radiation field.   }

\subsection{Behavior of the acoustic Resonant Drag Instability}\label{sub: RDI regimes}

As discussed in detail in \citet{Squire2018}, \hs, and \msh, the acoustic RDI 
exhibits different behaviors based on a dimensionless scale parameter $k \,c_{s} t_{s}$, where
$k$ is the wavenumber of the mode. 
In \cref{app: linear}, we cover in detail the linear 
behavior of the acoustic RDI with a spectrum of grain sizes, showing, as expected, that 
the same parameter (which now covers a range of values because of the range in $\ad$) has a similar influence on the 
instability. 
Specifically, three regimes emerge: the instability is in the low-$k$ regime if  $k\,c_{s} t_{s}\lesssim \mu$; the mid-$k$ regime if  $\mu\lesssim k\,c_{s} t_{s}\lesssim \mu^{-1}$; and the high-$k$ regime if $\mu^{-1}\lesssim k\,c_{s} t_{s}$.
With a spectrum of grains, there is ambiguity surrounding these delineations ($t_{s}$ and $\mu$ depend on $\ad$), 
but they are nonetheless useful for general understanding (see \cref{app: linear} for more precise estimates).
In the low-$k$ regime, the fastest growing modes are generally non-resonant and grow in
the direction aligned with the drift, involving strong perturbations to both the gas and dust densities. We show
in \cref{sub: low k linear} that such modes are generally agnostic to the presence of a spectrum of grain sizes or the drift regime (constant versus non-constant). 
In contrast, modes in the mid- and high-$k$ regimes are  fastest growing at a specific ``resonant'' mode direction, 
where the projection of the dust drift speed  onto that direction is equal to the sound speed. They have a much stronger dust-density response than gas-density  response. Such 
modes behave similarly in the constant-drift regime with a spectrum of grain sizes, but are significantly modified in the non-constant-drift regime
because each grain size resonates with a different mode angle (see \cref{app: grain indep accel}).

In addition to the linear behavior, the three regimes  control the {acoustic RDI's} nonlinear evolution (\msh).
While gas motions and dust clumping driven by the {acoustic RDI} in the low-$k$ regime broadly resemble standard
driven supersonic  turbulence (although there are distinct differences),  the mid- and high-$k$ regimes  are very  different, with the
resonant mode structure remaining clear well into the saturated state and  across
all scales. In larger-$\ws$ cases, this manifests itself through thin dust filaments, which ``draft'' on the 
nearby dust and never reach a saturated turbulent steady state. 
For subsonic drift, the linear and nonlinear behavior is most similar to the 
low-$k$ regime (indeed, the subsonic instability at mid- to high-$k$ is non-resonant 
and depends on details of the equation of state and drag law; \hs).

Finally, it is worth reiterating from previous works that the acoustic RDI generally has no preferred scale 
in any of the three regimes. Rather, modes at smaller scales grow faster, with the growth rate $\Im(\omega)$
scaling as $\Im(\omega)\sim k^{2/3}$, $\Im(\omega)\sim k^{1/2}$, and $\Im(\omega)\sim k^{1/3}$, in the 
low-, mid-, and high-$k$ regimes, respectively. Thus, simulations cannot be converged in the conventional sense, 
in that a higher-resolution simulation will resolve faster-growing modes (in the absence of a small-scale dissipative effect such as viscosity). However, as shown by \msh\ (appendix B3), the bulk properties of the saturated state are effectively resolution independent  once box-scale  modes saturate nonlinearly.

\subsection{Simulation design}\label{sub: simulation set up}

We use the code GIZMO,\footnote{A public version of the code, including all methods used in this paper, is available at \href{http://www.tapir.caltech.edu/~phopkins/Site/GIZMO.html}{http://www.tapir.caltech.edu/~phopkins/Site/GIZMO.html} }
which solves the fluid equations using the second-order Lagrangian ``Meshless Finite Volume'' (MFV) method \citep{Hopkins2015}. Dust is included via the super-particle approach \citep{YoudinA2007,Hopkins2016a},
using a random sampling of grain sizes $\ad$ across the full continuous distribution (i.e., we do not use a 
set number of grain-size bins). The backreaction force of the dust on the gas is computed using a standard momentum-conserving scheme \citep{YoudinA2007}, with details of the scheme and a variety of numerical tests described in appendix B of \msh\ (although \msh\ considers 
only a single grain size, there are no significant  numerical complications that arise
from the use of a spectrum of grains).

The range of available parameter space for simulations using a grain-size spectrum is
extensive, even without magnetization and grain charge: it includes the drift-velocity regimes (constant versus non-constant, supersonic versus 
subsonic, and mixtures of each), stopping-time distributions (which could straddle the different $k$ regimes),  the distribution of grain masses, and the total dust-to-gas-mass ratio. 
Because  of this, and motivated by the goal of better understanding RDI-turbulence physics rather than 
detailed matching of specific astrophysical situations, we choose
in this article to focus on the detailed understanding of just two sets of RDI parameters. 
We supplement this by comparing these directly to simulations without 
dust backreaction ($\mu_{0}=0$), where turbulence 
is driven by large-scale external forcing to have a similar  velocity dispersion. 
{The purpose of this comparison is two fold: firstly, and most importantly, 
it enables us to probe the physics of dust clumping in RDI-generated turbulence by direct comparison to the better-understood 
case of standard (Kolmogorov) turbulence, revealing clearly their most important differences.
Secondly, in application to AGB-star winds or AGN outflows, the forced-turbulence simulation 
could provide a reasonable model for dust clumping   if the RDI were not operating. Specifically, one might expect larger-scale global instabilities (e.g., 
 Rayleigh-Taylor-like instabilities; \citealp{Krumholz2012}) to drive turbulence, which would then, through a turbulent cascade, drive
 fluctuations on the small scales considered here. Thus, the simulations act as a benchmark for how such a dust-driven wind might 
 clump dust in the absence of RDIs (e.g., at extremely small dust-to-gas ratios), although we caution that the
 magnitude of the turbulent driving is not at all tuned to explore this in detail (rather it is tuned to address the first point and probe the physics). }

The overall approach complements and builds on that of \msh\ and \citet{Hopkins2020}, which 
surveyed a wide range of parameters to understand how the RDI behaves in different regimes. Specifically,
 the results of \msh\ tell us that the most interesting computationally accessible RDI behaviour -- i.e.,
 that which exhibits the most interesting differences compared to standard turbulence -- occurs in the ``mid-$k$''
 range. 
 As discussed above (\cref{sub: RDI regimes}) this regime is also expected to show 
 more interesting differences  between the grain-spectrum and single-grain RDIs, so is, overall, the
 most obvious candidate for further study.\footnote{Unfortunately, reaching the true high-$k$ regime
has proved to be computationally challenging, because the width of the resonant wavelengths become increasingly narrow at increasing $k$, necessitating excessively high resolution. \msh\ considered some cases
around the boundary between the mid- and high-$k$ regimes, which were  similar to the mid-$k$ cases (see their appendix A for more information).} 
 Thus we will ignore the low-$k$ regime and/or subsonic drift in this study. Although they are
 potentially astrophysically relevant in many situations \citep{Hopkins2021}, these regimes 
 can likely be mostly adequately understood using a combination
 of RDI-related understanding from \msh\  and \hs\, and theories of collisions/clustering in turbulence 
with a spectrum of sizes without dust backreaction \citep[e.g.,][]{Pan2014a,Pan2014b,Hopkins2016a,Mattsson2019,Li2020}.

Based on the discussion of the previous paragraph, we focus on four simulations with a grain-size spectrum covering a factor of 100 ($\amax=100\amin$), each in a cubic box 
of size $L^{3}$.  These are: 
\begin{description}
\item[\textbf{Constant-Drift RDI}] This simulation sets $\aext\propto1/\ad$ such that $\ws$ is independent of $\ad$ and  $t_{s}\propto \ad$. The acceleration and grain size range (see below) is chosen such that $\ws\approx1.5$, which 
is convenient because the resonant angle at which the modes grow fastest ($\cos\theta_{k}=1/\ws$) is neither
too oblique nor too parallel, as well as being astrophysically reasonable. The outer-scale normalized wavenumbers of grains range from $(2\pi/L)c_{s}t_{s}\approx0.005$ to $(2\pi/L)c_{s}t_{s}\approx0.5$. 
\item[\textbf{No-Backreaction Constant Drift}] This simulation has parameters (including $\aext$) that match exactly Constant Drift RDI, but with 
no dust backreaction, and thus no RDI. Instead, the turbulence is externally driven, with the amplitude of the forcing chosen so that the resulting turbulence has a velocity dispersion that matches (as closely as possible) the Constant Drift RDI simulation.
\item[\textbf{Non-Constant-Drift RDI}] This simulation sets $\aext$ constant such that $\ws$ and $t_{s}$ both depend
on $\ad$ ($Q_{\rm abs}\sim\ad/\lambda_{\rm rad}$). The acceleration and grain size range  is chosen such that $\ws\sim \ad^{1/2}$ ranges between $\ws\approx 0.73$ (for the smallest grains) and $\ws\approx 12.7$ (for the largest grains). The outer-scale normalized wavenumbers of grains range from $(2\pi/L)c_{s}t_{s}\approx0.003$ to $(2\pi/L)c_{s}t_{s}\approx0.05$ (note that $t_{s}\sim \ad^{1/2}$ for the larger/faster grains).
\item[\textbf{No-Backreaction Non-Constant Drift}] This simulation has parameters (including $\aext$)  that match exactly Non-Constant Drift RDI, but with no dust backreaction and using external driving to match the turbulence amplitude of Non-Constant Drift RDI.
\end{description}
Note that in each case, $(2\pi/L)c_{s}t_{s}<1$ is chosen so that smaller-scale modes in the box do not move too 
far into the high-$k$ regime, where they may be adversely affected by numerical resolution. In the non-constant drift cases, the range in $\ws$ is chosen so
that most grains are supersonic, but the largest grains are not highly supersonic ($\ws$ not too much larger than $\simeq 10$), which was shown 
by \msh\ to cause particularly extreme behavior (drafting) that does not necessarily converge in time (owing in part due to the finite periodic box assumption that we adopt). The equilibrium $\ws$ and $(2\pi/L)c_{s}t_{s}$ are shown
below in the insets of \cref{fig:time.evolution.sda,{fig:time.evolution.ca}}.

{\subsubsection{Numerical parameters}}

{In each case, it is convenient to use code units defined by $\langle \rho_{g}\rangle=1$, $c_{s}=1$, $\bar{\rho}_{d,{\rm int}}=1$, and $L=1$. Then, 
$\ws(\ad)$ and $t_{s}(\ad)$ are numerically determined by the combination of $\ad$ and $\aext$ (in code units); in order to obtain 
the simulation parameters discussed above, we need 
$\amax=0.15=100\amin$ with $\aext = 3/\ad$ for the constant-drift simulations, and  $\amax=0.0878=100\amin$ with $\aext =1400$ for the non-constant-draft simulations.}
Grains are initialized with their equilibrium drift velocity ($\ws(\ad)/t_{s}(\ad)=\aext$), which is also 
present in the driven cases without backreaction (unlike previous studies of turbulent grain dynamics). {We use periodic boundary conditions with a resolution of 
$256^{3}$ gas particles and $4\times256^{3}$ dust particles in all simulations}. This was chosen based on the scaling tests in \msh\ (appendix B), which showed  reasonable convergence in the mid-$k$ regime between $128^{3}$ and $256^{3}$
 (although RDI simulations of this type can never be truly converged because the instability generally grows fastest at the smallest scales of the box).\footnote{{The numerical time step in all simulations is strongly limited by the 
integration of the smallest grains, which have a very short stopping time. This increases their computational cost by a factor of ${\simeq}10$ compared
to an equivalent system without dust, which, combined 
with the relatively slow saturation of the RDI (up to $t\simeq 30 L/c_{s}$),  makes the simulations relatively computationally expensive despite their modest resolution.}} \msh\ also showed  convergence in RDI dust probability-density functions (PDFs) once the dust resolution approached that of the gas (a factor $4$ lower gives good results, but not a factor 16 lower) in line with previous works \citep[e.g.,][]{Bai2010b}; our choice of $4\times256^{3}$ dust particles is made due to the wide range of grain sizes in our simulations and is explored further in  \cref{app: linear gizmo comparison}.
 Although we use an MRN  dust-mass distribution in all cases $d\mu/d\ln\ad\propto \ad^{1/2}$, we
use an equal number of numerical super-particles randomly sampled across each logarithmic size range (i.e., the super-particles change in their ``total'' super-particle mass proportionally to $\ad^{1/2}$)
in order to not under-resolve the small particles. 
\revchng{We note that some of the
convergence problems that are well known in the numerical computation of polydisperse dust instabilities 
seem to be  less severe in our simulations than in previous works on the protoplanetary disk streaming instability \citep{Krapp2019,Paardekooper2021}.
The difference here may be due to differences between the linear properties of the acoustic RDI and the streaming instability,
or could relate to the   numerical method, whereby grains are randomly sampled in size space across the full distribution, each with their own stopping time (it is thus effectively a type of Monte-Carlo 
integration scheme). We investigate this convergence further in \cref{app: linear gizmo comparison} by comparing the early phases of GIZMO simulations to linear results;
however,  in order to understand the  influence of both the numerical method and the instability's properties, further study of these issues would be beneficial.  }

In the no-backreaction, driven cases, we use time-correlated incompressible (solenoidal) forcing at the  largest scales (Fourier modes with $k\leq4\pi/L$; \citealp{Hopkins2016a}). The  correlation time is $0.5L/c_{s}$ in the constant-drift simulation and $0.2L/c_{s}$ in the non-constant-drift simulation, because the turbulence in the non-constant-drift case 
has higher Mach number, so a shorter box-crossing time.
We use the total dust-to-gas-mass ratio $\mu_{0}=0.01$ for both of the RDI simulations ($\mu_{0}=0$ for the no-backreaction simulations).

\subsubsection{Grain-size bins for diagnostics}\label{sub: dust binning}

Although our simulations involve a continuous distribution of grain sizes, for most diagnostics -- for instance, 
any quantity involving dust density or velocity as per \cref{eq: vd rhod defs} -- it is necessary
to bin the dust size distribution. We choose to do this across $8$ logarithmically spaced bins, which we label 
with $\ab{1}$ for the smallest grains, through to $\ab{8}$ for the largest grains. More precisely, the label $\ab{i}$ 
refers to grains with $\ad$ between $10^{b_{i}}\amax$ and $10^{b_{i+1}}\amax$, where $b_{i}=(-2,-1.75,-1.5,-1.25,-1,-0.75,-0.5,-0.25,0)$. So, for example, grains of size $\ab{1}$  have a density and bulk velocity  given 
by \cref{eq: vd rhod defs}, with $\abc{l}=10^{-2}\amax=\amin$ and $\abc{h}=10^{-1.75}\amax\approx 1.78 \amin$.

\subsection{Mapping to astrophysical applications}\label{sub: mapping to astro}

Our simulations are not intended to map to a specific astrophysical object, but
rather study the generic behavior of RDI-generated turbulence. Here, we 
outline how the simulation units -- with $\langle \rho_{g}\rangle=1$, $c_{s}=1$, $\bar{\rho}_{d,{\rm int}}=1$, and $L=1$ (the box size) -- translate into various astrophysical situations and processes of interest. The 
two important properties are  the grains' drift velocity -- which depends on the radiation field, grain size, and other 
gas properties -- and the physical box scale $L_{\rm phys}$, which depends on the gas density and grain sizes.
Following \hs, we  briefly consider as examples asymptotic giant-branch (AGB) stars, Active Galactic Nucleii (AGN), and star-forming regions (giant molecular clouds; GMCs),
which are the contexts most relevant to uncharged dust; more 
detailed estimates for a wider range of situations are given in \citet{Hopkins2018a} (see their figure 6).

A simulation can be mapped to a given physical situation by comparing the box size $L=1$ to the grain-drift length at $\ws=0$, $L_{\rm drift}=c_{s} t_{s}|_{\ws=0} = \sqrt{\pi/8}\ad \bar{\rho}_{d,{\rm int}}/\rho_{g}$. In order
to obtain the desired $\ws$ and $(2\pi/L)c_{s}t_{s}$, as discussed above, the
constant-drift simulations use $\amax=0.15=100\amin$, while the non-constant-drift simulations use $\amax=0.0878=100\amin$. Thus, the box scale translates into $L_{\rm phys}\approx6.7\, \amax \bar{\rho}_{d,{\rm int}}/\rho_{g}$ and $L_{\rm phys}=11.4\, \amax  \bar{\rho}_{d,{\rm int}}/\rho_{g}$ for the constant- and non-constant-drift 
cases, respectively.
\begin{description}
\item[\textbf{Star-forming regions}]
\hs\ estimates GMC conditions  under reasonable assumptions (e.g., a cloud of $\sim\!10{\rm pc}$ that has converted $\sim\!0.1$ of its mass into stars), finding 
$\ws\sim10 $ for larger grains with $Q_{\rm abs}\sim 1$ ($\ws$ is larger closer to a more luminous source and at lower gas density). Depending on the wavelength of the radiation and the dominant grain sizes, this is  in reasonable agreement to both the constant- and non-constant-drift simulations. The above estimates yield $L_{\rm phys}\simeq 20 {\rm pc}$, showing that the smaller-scale, shorter-time dynamics in our simulations will 
apply to larger scales around a GMC. 
\item[\textbf{AGB-star winds}]
The envelopes/winds of AGB stars are dust laden and driven by radiation pressure. Using similar estimates to 
\hs\ (a stellar luminosity of $\sim\!10^{5}M_{\sun}$ with  mass-loss rate $\sim\!10^{-4}M_{\sun}{\rm yr}^{-1}$) yields
$\ws \sim 0.1\rightarrow 10$, which is in  the range probed by either simulation. Similarly, estimating the gas density at $\sim\!1{\rm AU}$ from a star with a wind velocity $\sim\!10{\rm kms}^{-1}$ yields $L_{\rm phys}\sim 5\times10^{-5}{\rm AU}$ showing that our boxes are probing relatively small scales inside the outflowing wind, where the RDI would seed small-scale clumping.
\item[\textbf{AGN-driven outflows}]
\hs\ and \citet{Hopkins2018a} estimate very fast drift velocities, from $\ws\sim 1$ for small grains to $\ws\gtrsim 100 $ for larger grains, in the ``dusty torus'' region around an AGN with luminosity $\sim\!10^{46}{\rm erg}{\rm s}^{-1}$. Using $n\sim 10^{6}{\rm cm}^{-3}$, as appropriate to a   denser closer-in  region, yields $L_{\rm phys}\sim 2\times 10^{-4}{\rm pc}$, showing that (as for the AGB wind), the numerical box represents a small patch of the larger system, representing fast dust dynamics on small scales.
\end{description}
Overall, we see that our  simulation parameters are  likely most  applicable to  AGB winds; GMC-related  applications would be better served by a somewhat smaller-scale box, while the drift velocities in the AGN are somewhat more extreme than simulated. {Also, grain charging and MHD effects are likely 
more important for many GMC-like conditions, which would change our results substantially \citep{Hopkins2021}.} However, parameters in all three cases vary enormously, and the physical ideas we explore are generally relevant to different
regions of all three cases. Our chosen dust-to-gas-mass ratio of $\mu_{0}=0.01$  
broadly applies to most situations, although this can vary significantly both above and below this value (see, e.g., \citealt{Knapp1985,Dharmawardena2018,Wallstroem2019} for some examples of AGB winds).

\begin{figure*}
\begin{center}
\includegraphics[width=1\textwidth]{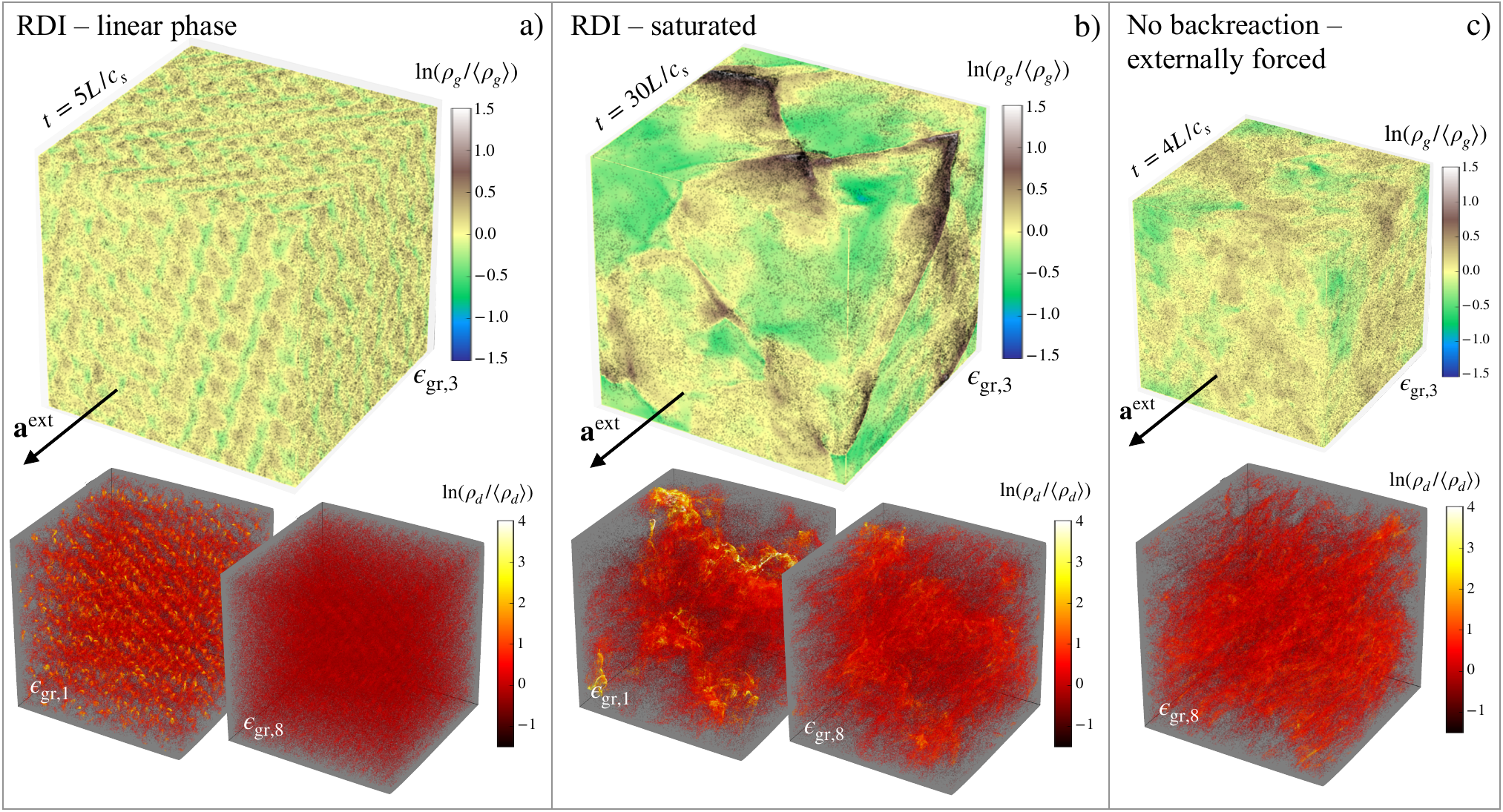}
\caption{Dust and gas visualizations for the constant-drift simulations. Top images show the gas density on the surface of the cube with medium-sized ($\ab{3}$) grains shown as black dots, while bottom images show volume renderings of the dust density for small ($\ab{1}$) and large ($\ab{8}$) grains (the dust binning is described in   \cref{sub: dust binning}).  Panel (a) shows the near-linear phase of the RDI at early times and  panel (b) shows show its saturated turbulent state at late times (see also \cref{fig:time.evolution.sda}). Panel (c)  illustrates the saturated turbulence in the no-backreaction constant-drift simulation, where the turbulence is externally forced (only large grains are shown in the bottom image). 
}
\label{fig:panels.sda}
\end{center}
\end{figure*}
\begin{figure*}
\begin{center}
\includegraphics[height=\tfigh]{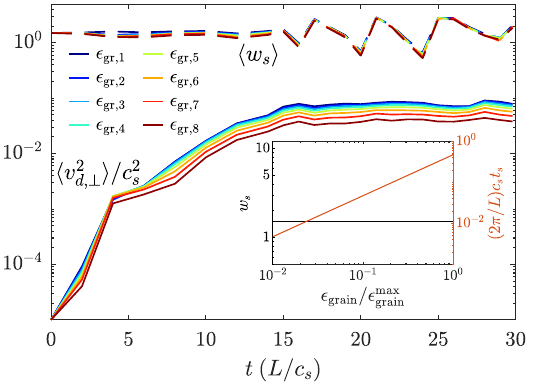}~\includegraphics[height=\tfigh]{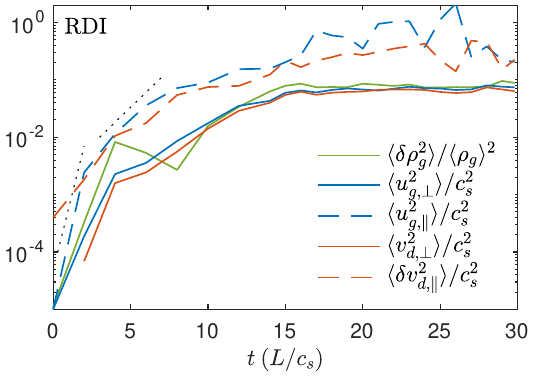}\includegraphics[height=\tfigh]{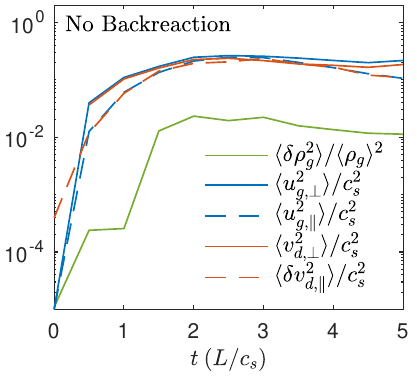}
\caption{Time evolution of bulk quantities in the constant-drift simulations. The left-hand panel shows the evolution
of dust-related quantities in the RDI simulation for different-sized grains from small ($\ab{1}$; blue) to large ($\ab{8}$; red), showing the measured perpendicular dust velocity dispersion $\langle \delta v_{d,\perp}\rangle= \langle v_{d,\perp}^{2}\rangle$ with solid lines, and the drift velocity $\langle \ws\rangle=\langle v_{d,\|}\rangle$ with dashed lines.  The inset shows the
equilibrium $\ws(\ad)$ (black, left axis) and $t_{s}(\ad)$ (red, right axis). Similar quantities for the gas and dust averaged over all bins
are shown in the middle panel for the RDI simulation, and in the right-hand panel the no-backreaction case with externally driven turbulence. Their comparison some clear differences in the structure of RDI-driven turbulence (larger parallel velocity fluctuations and stronger relative density fluctuations) that cannot be matched by isotropic incompressible driving. In taking averages, gas quantities are weighted by the gas mass and dust  quantities by the dust mass.
The black dotted lines in the middle panel show the linear growth rates of resonant modes with $k\approx 32\pi/L$ (approximately $1/16$ of the box scale) and $k\approx 2\pi/L$ (approximately the box scale); see   \cref{eq: size dep final ans} and \cref{fig:linear adep}.
}
\label{fig:time.evolution.sda}
\end{center}
\end{figure*}

\begin{figure*}
\begin{center}
\includegraphics[width=1\textwidth]{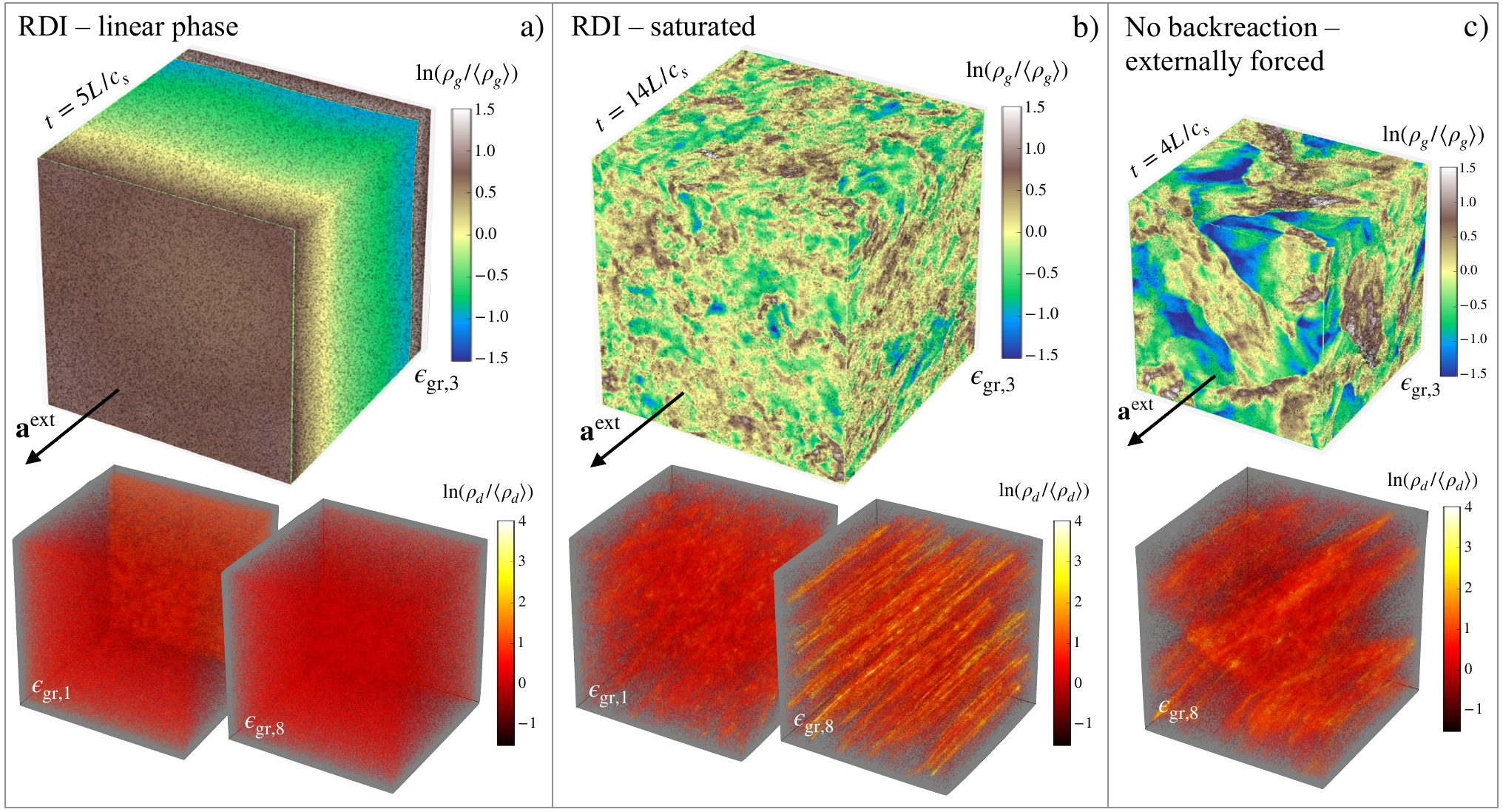}
\caption{As for \cref{fig:panels.sda}, but for the non-constant-drift simulations. 
}
\label{fig:panels.ca}
\end{center}
\end{figure*}

\begin{figure*}
\begin{center}
\includegraphics[height=\tfigh]{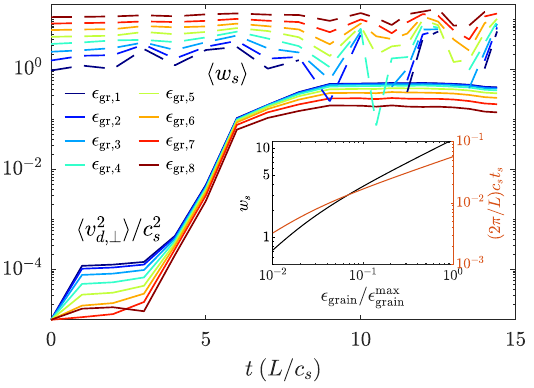}~\includegraphics[height=\tfigh]{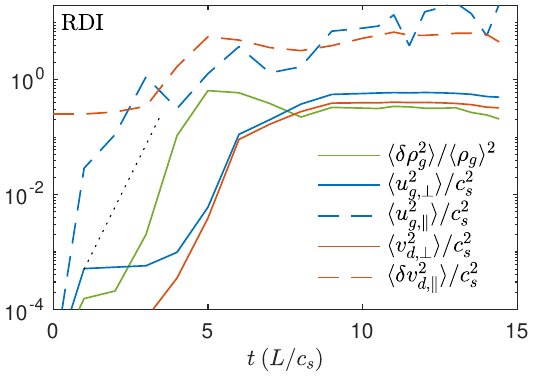}\includegraphics[height=\tfigh]{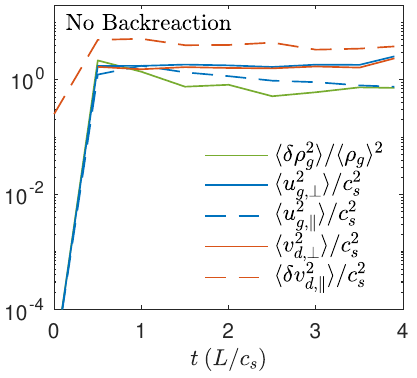}
\caption{As for \cref{fig:time.evolution.sda}, but for the non-constant drift simulation. Because the larger grains, with most of the mass,
move faster  in this case, the resulting turbulence is trans-sonic, with $\langle u_{g}^{2}\rangle\sim c_{s}^{2}$
and larger density fluctuations than the constant-drift simulation. The black dotted line in the middle panel indicates the linear growth 
rate of {the box-scale mode seen in the left panels of \cref{fig:panels.ca}; the details and convergence of this linear growth phase are assessed 
in detail in \cref{app: linear gizmo comparison}}.
}
\label{fig:time.evolution.ca}
\end{center}
\end{figure*}

\section{Results}\label{sec: results}

In this section we present results of the four GIZMO simulations outlined above (\cref{sub: simulation set up}). We 
start with a general exploration of the time evolution and turbulence  structure (\cref{sub: results: time evolution,sub: results: statistics}) then consider more detailed statistics related to dust clumping and collisions in \cref{sub: results: collisions}.

\subsection{General morphology and time evolution}\label{sub: results: time evolution}


In \cref{fig:panels.sda,fig:time.evolution.sda} we show  three-dimensional visualizations of the 
turbulence structure and the time evolution of important quantities for the constant-drift simulations. The 
same is shown for the non-constant-drift simulations in \cref{fig:panels.ca,fig:time.evolution.ca}. 
In each case, the turbulence structure is illustrated  during  the instability-growth phase (left-hand panels)  and once it saturates nonlinearly (middle panels)  for the RDI simulations, 
 and compared to the gas and dust structure of the externally forced runs without dust backreaction (right-hand panels). 
 The bottom panels visualize the smallest and largest grains in the RDI, to compare differences in their structure.
 The time-evolution panels (\cref{fig:time.evolution.sda,fig:time.evolution.ca}) show how  
 velocity dispersions and drift velocities vary with grain size (left-hand panels; insets show the equilibrium dust parameters) and gas and dust 
 velocity dispersions integrated over all grain sizes (middle and right-hand panels), comparing 
 the RDI cases to the driven-turbulence ones (right-hand panels). 
 These plots illustrate the basic time evolution of the RDI  through the linear phase and  saturation, and allow 
 simple comparison to the saturated state of driven turbulence.

The early-time growth in the constant- and non-constant-drift cases is rather different and well explained 
by the linear mode structure in each case. In the constant-drift case, the growth 
is initially rapid but slows in time, broadly matching the predicted linear growth rates  (dotted lines in the middle panel of \cref{fig:time.evolution.sda}), which increase monotonically with scale (see  \cref{fig:linear adep}).
This behavior is expected and discussed in previous works on RDI evolution (\citealp{Seligman2019}; \msh); it results from 
smaller-scale modes growing and saturating nonlinearly more rapidly than large-scale modes, so that the growth of a bulk quantity
(such as $\langle u_{g}^{2}\rangle$) is first dominated by the faster smaller scales, then, at later times (once the small-scales saturate), by the slower
larger scales. Such evolution is also clearly seen in the morphology in \cref{fig:panels.sda}: there is strong clumping of small grains at
small scales by $t\approx5 L/c_{s}$ even though the turbulence is far from full saturation at this point. The 
structures in the gas and dust, which clearly show the RDI resonant angle ($\theta_{k}=\cos^{-1}(\ws^{-1})\approx 48^{\circ}$)
are broadly similar to those that develop in saturation at the box scale (middle panel). We see that the smallest grains 
exhibit the strongest clumping (note higher-density patches in the bottom panels)  and have a modestly higher velocity dispersion in the
saturated state. However, it is also clear that large and small grains are undergoing similar dynamics:
 $w_{s}$ remains remarkably similar for all grains even as it fluctuates in time (left-hand panel of \cref{fig:time.evolution.sda}), and 
high-density  regions of large and small grains are clearly correlated spatially (bottom middle panels of \cref{fig:panels.sda}). Thus, as suggested by linear calculations (\cref{app: grain dep accel}), 
 the constant-drift RDI  involves different grain sizes interacting with the gas in similar ways, 
 driving resonant modes that cause strong clumping for all sizes concurrently.

The non-constant-drift RDI is more complex, with significant differences between the dynamics
of small and large grains. As discussed in detail in \cref{app: grain indep accel},
a number of  different linear-instability mechanisms can operate and/or dominate in the non-constant-drift RDI, and this 
is also true for the chosen scale and parameters of the simulation. In particular, there exists a large-scale parallel 
mode that resembles a backward-propagating sound wave, which has a similar growth rate to a smaller-scale, more oblique resonant mode 
that predominantly effects the largest grains (see \cref{fig:linear aindep}).
{In \cref{app: linear gizmo comparison}, we test the detailed linear growth of these modes across different scales, 
showing generally good agreement with linear predictions (see \cref{fig: nl lin comparison large,fig: nl lin comparison k theta}).}
The {effect} of this linear mode  structure can be seen in the turbulence morphology that develops, as well as 
in the time evolution of the RDI. We see the formation of a box-scale strong shock at early times (left-hand panel of \cref{fig:panels.ca}), which
creates a strong  density contrast in the gas and the smaller dust grains (because they are better
coupled to the gas).  It is also clear in the gas and dust  time evolution (middle panel of \cref{fig:time.evolution.ca}), 
manifesting  as a large density dispersion that develops until $t\approx 5L/c_{s}$, at which point in breaks up and grows  larger
turbulent velocity fluctuations into the true steady state for $t\gtrsim 8L/c_{s}$.
The visual turbulent morphology in the saturated state at late times (middle panel of \cref{fig:panels.ca}) is quite different from the driven 
turbulence  (right-hand panel), with
smaller-scale structures in the gas caused by highly elongated dust velocity filaments. We interpret this behavior as
being due to the quasi-resonant mode, which affects only the largest grains and is highly oblique because of their fast drift velocity  (\cref{fig:linear aindep} inset {and \cref{fig: nl lin comparison k theta}}); indeed, there is strong, oblique clumping of large grains but not small grains (middle-lower panel).
Further evidence for the different 
dynamics of small and large grains is seen in the evolution of $\ws$ (left-hand panel of \cref{fig:time.evolution.ca}),
which exhibits much larger fluctuations for small grains than large grains in the saturated turbulence, even 
though their perpendicular dust velocity dispersions (solid lines) are similar. 

\begin{figure}
\begin{center}
\includegraphics[width=0.9\columnwidth]{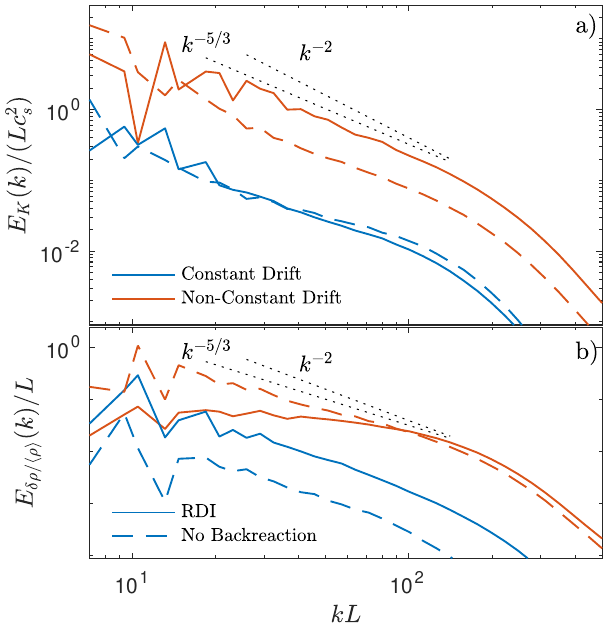}
\caption{Top panel: gas velocity spectra from all four simulations, with blue lines showing the constant-drift simulations, red lines showing the non-constant-drift simulations, solid lines showing RDI simulations, and dashed lines showing no-backreaction (externally forced) cases. We see similar behavior in all cases, with kinetic energy spectra $\sim\!k^{-5/3}$ that broadly match
previous results for subsonic and trans-sonic turbulence. Bottom panel:  gas density spectra, which also broadly match expectations, with a significantly flatter spectrum in the higher-Mach-number non-constant-drift case.
}
\label{fig: spectra} 
\end{center}
\end{figure}

\subsubsection{The level of turbulent driving}\label{subsub: turb driving level}

It is worth briefly commenting on some differences between RDI turbulence and the externally forced runs without dust backreaction.
The level of forcing in the no-backreaction runs was chosen to match the RDI saturation as best as possible; however,
 the choice of isotropic forcing in the driven runs means that
$u_{g,\perp}^{\rm rms}\approx 2 u_{g,\|}^{\rm rms}$ (where $u_{g,\cdot}^{\rm rms}= \langle u_{g,\cdot}^{2}\rangle^{1/2}$ is the root-mean-squared gas velocity), while the saturated state of the RDI runs are distinctly non-isotropic, with much larger parallel velocity dispersions
$u_{g,\|}^{\rm rms}\gg u_{g,\perp}^{\rm rms}/2$. Thus, although both RDI cases have a modestly larger total velocity dispersion than 
their equivalent no-backreaction runs ($u_{g}^{\rm rms}\approx 0.9c_{s}$ versus $u_{g}^{\rm rms}\approx 0.6c_{s}$ for 
the constant-drift case; $u_{g}^{\rm rms}\approx 3.6c_{s}$ versus $u_{g}^{\rm rms}\approx 1.7c_{s}$ for 
the non-constant-drift case), their perpendicular velocity dispersions are smaller ($u_{g,\perp}^{\rm rms}\approx 0.3c_{s}$ versus $u_{g,\perp}^{\rm rms}\approx 0.5c_{s}$ for 
the constant-drift case; $u_{g,\perp}^{\rm rms}\approx 0.8c_{s}$ versus $u_{g,\perp}^{\rm rms}\approx 1.4c_{s}$ for 
the non-constant-drift case). In addition, we see that externally forced turbulence has a somewhat lower density dispersion
than RDI turbulence in the constant-drift case, while the opposite is true for the non-constant-drift simulations. 
While it might be possible to rectify some of these discrepancies using non-isotropic driving, such an 
exercise would not necessarily be helpful: the differences arise from fundamental differences in the structure and evolution 
of RDI turbulence and driven turbulence, which is exactly the physics that we wish to study. However, it is
important to keep them in mind as we explore some of the differences in more detail. 

\subsubsection{Turbulent Stokes numbers of grains}\label{subsub: stokes numbers}
A parameter of particular importance for turbulent grain dynamics is the ratio of stopping time 
to the  turnover time of the turbulence, known as the Stokes number, ${\rm St}$. For comparison to previous 
results  on dust dynamics in turbulence without backreaction  \citep[e.g.,][]{Ormel2007,Pan2013,Pumir2016,Mattsson2019} it is helpful to estimate ${\rm St}$ at the largest and smallest resolvable scales in our 
simulations. The large-scale turnover time is estimated as $\tau_{L}\sim L/u_g^{\rm rms}$, which (using 
the externally forced turbulence values from above to avoid dealing with anisotropic turbulence) is $\tau_{L}\simeq 1.7 L/c_{s}$ for the 
constant-drift case, and $\tau_{L}\simeq 0.6 L/c_{s}$  for the non-constant-drift case. As expected, the saturation of the turbulent runs, seen in the right-rand panels of \cref{fig:time.evolution.sda,fig:time.evolution.ca}, occurs over a timescale $\sim\!\tau_{L}$.
The fastest time scale in the turbulence, which occurs at small scales and is termed $\tau_{\nu}$,
 can be estimated from basic Kolmogorov arguments. For a $\sim\!k^{-5/3}$ velocity spectrum, which is at least roughly 
 valid  for the trans-sonic turbulence here (see \cref{fig: spectra}), the turnover time of structures of lengthscale $l$ scales as $l^{2/3}$. Using twice the average point spacing at our fiducial resolution, $l_{\nu}\simeq L/128$, 
 as an estimate for the smallest scales available before numerical viscosity 
 starts damping fluid motions, we estimate $\tau_{\nu}\simeq 0.07L/c_{s}$ and $\tau_{\nu}\simeq 0.02L/c_{s}$ for the constant- and non-constant-drift cases, respectively. 
Denoting the outer scale and smallest-scale Stokes numbers as ${\rm St}_{L}=t_{s}/\tau_{L}$ and  ${\rm St}_{\nu}=t_{s}/\tau_{\nu}$, respectively, we thus find
\begin{gather}
{\rm St}_{L}\simeq 0.01\left(\frac{\ad}{\amax}\right)^{1/2},\: {\rm St}_{\nu}\simeq 0.4\left(\frac{\ad}{\amax}\right)^{1/2},\quad \text{Non-constant drift},\nonumber\\
{\rm St}_{L}\simeq 0.05\frac{\ad}{\amax},\: {\rm St}_{\nu}\simeq 1.1\frac{\ad}{\amax},\quad \text{Constant drift}.
\label{eq: stokes in simulations}
\end{gather}
We see that these simple estimates would naively suggest that nearly all grains in our simulations are  ``well coupled'' (${\rm St}\lesssim1$) to the gas turbulence across all scales, meaning that they should passively trace gas motions. We will show below that small grains in the constant-drift RDI are actually not ``well coupled'' to the gas turbulence in this sense, despite the fact that 
${\rm St}_{\nu}\simeq 0.01$ for this population. This is likely because the compressive and dust-drift motions associated with the instability are both faster (thus increasing ${\rm St}$)
and more effective at driving dust clumping. 
Finally, it is worth noting that the RDI quasi-linear saturation estimate of  \msh, $u_{g}^{\rm rms}\sim \mu^{1/2} (c_{s}t_{s}/L)^{-1/2}$, which compared well 
against single-grain-size simulations, yields similar estimates for ${\rm St}$; ${\rm St}_{\nu}\sim \mu^{1/2}(c_{s}t_{s}/L)^{1/2}(l_{\nu}/L)^{-2/3}$, for the Stokes number on scale $l_{\nu}$, which is generally $\ll1$ for outer scales in the low- or mid-$k$ regime with $\mu\ll1$, except at very small scales. 
Thus, it seems that simple (Kolmogorov) estimates 
of turbulent Stokes numbers in RDI turbulence generically yield  rather small values (${\rm St}\ll1$) compared to what 
is usually  necessary to strongly clump grains in turbulent flows (i.e., ${\rm St}\gtrsim 1$; see, e.g., \citet{Hopkins2016a}, who see only very weak clumping
for ${\rm St}\ll 1$ in compressible turbulence with the same numerical methods).

{The discussion above also suggests that another clumping regime could emerge at extremely small scales: at scales where ${\rm St}_{\nu}\gg 1$,
the turbulent cascade may start to dominate the clumping for grains that  are not strongly affected directly by the RDI. 
Given the efficiency of the constant-drift RDI at clumping all grain sizes, such an effect would likely be important only for smaller grains in
the non-constant-drift regime, and would manifest as an enhanced clumping at larger resolution. Using 
the estimate \eqref{eq: stokes in simulations}, we see that accessing this regime --  ${\rm St}_{\nu}\gtrsim 1$ for the smallest grains -- requires
${\rm St}_{\nu}$ to be ${\simeq100}$ times larger than the simulations presented here, thus requiring ${\simeq}1000$ times more resolution
elements in each direction (assuming $\tau_{\nu}\propto l^{2/3}$). This is clearly unattainable with present resources, although it is possible
that a similar regime could be accessed with a different set up.}

\subsection{The turbulent steady state}\label{sub: results: statistics}

In this and the following section, we explore more detailed statistics of the saturated states of RDI turbulence, comparing
to the turbulence runs as a reference. 

\paragraph*{Spectra}
The simplest measure of the scale-dependent turbulent structure is the spectrum, which 
is illustrated for all four cases in \cref{fig: spectra}. The kinetic energy spectra (panel a) of RDI turbulence and externally forced (no-backreaction) turbulence are
seen to be quite similar.  Spectral slopes in the inertial range ($20/L\lesssim k\lesssim 200/L$) 
are approximately $\sim\!k^{-5/3}$, as expected for a standard Kolmogorov cascade, and are slightly steeper in the 
non-constant-drift runs, as expected because of their higher Mach numbers (velocity spectra steepen to $\sim\!k^{-2}$ for highly supersonic flows; e.g., \citealt{Federrath2013}). 
The consistency of the RDI and forced velocity spectra is interesting, given the clear differences in their structures observed in \cref{fig:panels.sda,fig:panels.ca}. For example, we see that although the non-constant-drift RDI turbulence  looks quite 
different to its  forced counterpart, with smaller-scale features, the scaling of the spectra at smaller wavenumbers $k\gtrsim 20/L$
 is very similar. The density spectra (panel b) are also consistent with previous works \citep{Konstandin2016}, 
 although we see more interesting differences between the RDI and externally forced turbulence runs. 
The constant-drift runs exhibit similar, relatively steep ($\sim \!k^{-2}$) spectral scaling, but 
with stronger compressive motions in the RDI (see also \cref{fig:time.evolution.sda}); however, the non-constant-drift density spectra
are quite different, perhaps because of the higher-mach-number  parallel flows in the RDI case. 

\begin{figure}
\begin{center}
\includegraphics[width=0.9\columnwidth]{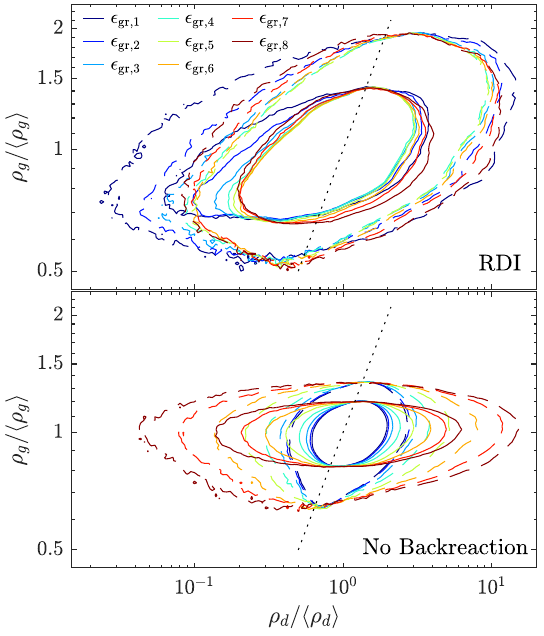}
\caption{Two-dimensional volume-weighted  PDF of the local (resolution-scale) dust density $\rho_{d}(\ad)$ and gas density $\rho_{g}$ for the constant-drift simulations. Solid lines show 
contours at $0.1$ of the maximum, while dashed lines show the contour of $0.01$ of the maximum, with 
each color showing the PDF for different grain-size bins from the smallest grains $\ab{1}$ in blue, to the the
largest $\ab{8}$ in dark red. The PDFs are averaged over 3 snapshots during the saturated phase.  The two cases are markedly different: in  externally forced turbulence case without backreaction (bottom panel)
the smallest grains mostly trace the gas and larger grains exhibit more clumping; in the RDI turbulence, the smallest
grains instead exhibit the strongest clumping, with a long tail extending to very small dust densities. 
}
\label{fig:2d hist sda}
\end{center}
\end{figure}

\begin{figure}
\begin{center}
\includegraphics[width=0.9\columnwidth]{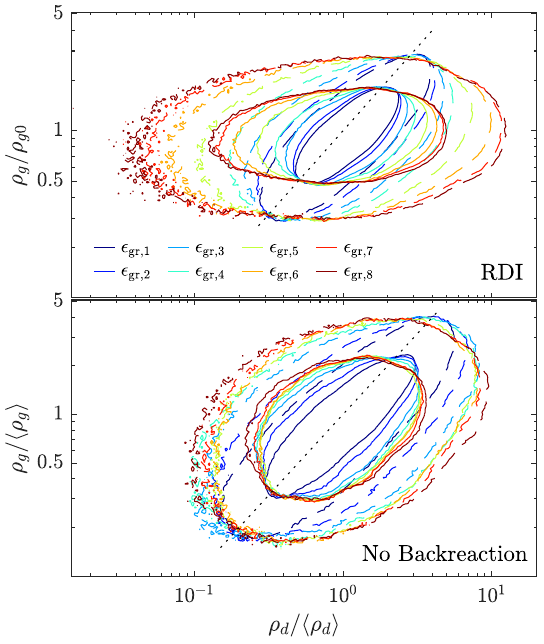}
\caption{Same as \cref{fig:2d hist sda}, but for the non-constant-drift simulation. The RDI PDF is  more
similar to externally forced turbulence in this case, although we still see significantly wider dust-density distributions (despite the 
somewhat narrower gas-density distribution), particularly for the largest grains that are more strongly affected 
by the mid-$k$ RDI (see text).
}
\label{fig:2d hist ca}
\end{center}
\end{figure}

\paragraph*{Dust and gas distribution}
A useful measure of the level and structure of 
dust clumping is the joint Probability Density Functions (PDFs) of dust and gas density, the contours of which
 are shown in \cref{fig:2d hist sda,fig:2d hist ca} for the 
constant- and non-constant-drift cases, respectively. 
These illustrate how regions of high gas density correlate with those of high dust density, and likewise for low-density regions. The black dotted lines illustrate the one-to-one correlation 
that would be observed if dust were perfectly coupled to the gas.

Let us first consider the constant-drift case (\cref{fig:2d hist sda}), which shows a significant difference between the RDI-generated turbulence (top panel) 
and the  forced turbulence without dust backreaction (bottom panel). This is surprising, given 
the similarity of their spectra. Most clearly, we see that in RDI turbulence the smallest grains (blue contours) 
exhibit larger fluctuations than the larger grains (red contours), particularly at low-densities, in stark contrast to the no-backreaction runs. The characteristic shape -- with a high-probability of low dust density in lower-gas-density regions -- was 
also seen in \msh\ and seems to be a typical  feature of the saturated state of the mid-$k$ (or high-$k$)  RDI. 
The much wider dust density distribution of larger grains in externally forced turbulence is well explained by their relative turbulent Stokes
numbers (which are quite small across all scales for the smallest grains; see \cref{subsub: stokes numbers}, \citealp{Pan2013,Hopkins2016a,Mattsson2019}), so the
fact that this is not true for the RDI turbulence is an important illustration of its different grain-clumping mechanisms.

In contrast, the non-constant-drift RDI PDFs appear more similar to externally forced turbulence (\cref{fig:2d hist ca}), although 
the dust-density distribution is wider for the larger grains in the RDI case (i.e., the difference between the large and small grains is larger). This is likely because  the highly oblique resonant instability, 
which is driven  only by the large grains (see middle panel of \cref{fig:panels.ca}), is particularly efficient at grain
clumping; the backward-propagating sound-wave mode, in contrast, clumps grains in a similar
way to standard  turbulence, creating a similar (small) density dispersion in the smaller grains.
This explanation is commensurate with the  enhanced clumping of all grains (compared to driven turbulence) 
in the constant-drift case, where the resonant instability operates with all grain sizes.

\subsection{Dust clumping and collisions}\label{sub: results: collisions}


The observations above, along with previous results (\msh; \citealp{Seligman2019,Hopkins2020}), show that the grain-clumping  
and its dependence on grain size can be very different in the RDI compared to externally forced turbulence without backreaction. 
The most obvious question that arises is whether 
these differences significantly change grain-collision statistics in the RDI compared to previous theories \citep{Voelk1980,Zaichik2006,Pan2013,Mattsson2019,Li2020}. 
Two key properties influence grain collisions and are needed to compute the collision kernel: the first is the relative
clumping of grains in space \citep{Maxey1987},  the second is the relative  velocity of grains that do collide \citep[e.g.,][]{Pumir2016}. Put together, 
these can be used to construct estimates for the collision rate and the sticking-bouncing-fragmentation probabilities \citep[e.g.][]{Garaud2013,Pan2014a}, an understanding of which is key for estimating how turbulence influences grain growth  
 in astrophysical scenarios (e.g., in an AGB outflow).
A full, careful estimate of these probabilities requires grappling 
 with a number of subtle convergence issues; for instance, relative grain velocities depend on particle separation, 
 and estimating the true collision velocity (at near-zero particle separation) requires a careful consideration   of 
 how different physical effects contribute to the relative velocity \citep{Falkovich2002}
 and how these are affected by numerics \citep{Pan2014b,Haugen2021}. Our goal here 
is less detailed -- to compare and contrast the relative clumping and grain-collision velocities
between the constant- and non-constant-drift RDIs and externally forced turbulence without dust backreaction. 
{We find much stronger clumping and reduced collision velocities 
of grains in the constant-drift RDI, with qualitatively different trends for small grains and grains of different sizes.
These trends are sufficiently strong to reveal clear differences between grain-grain collisions in the RDI and externally forced turbulence. We thus
forgo a  
a careful study of the convergence to the zero-particle-separation limit, which would be a formidable task for RDI turbulence 
given the wide range of different regimes \citep{Hopkins2018}.  
Overall, our results suggest}
that the RDI could strongly enhance grain growth rates  in outflows, especially in the constant-drift regime ($\ad\gg \lambda_{\rm rad}$).

\begin{figure}
\begin{center}
\includegraphics[width=0.88\columnwidth]{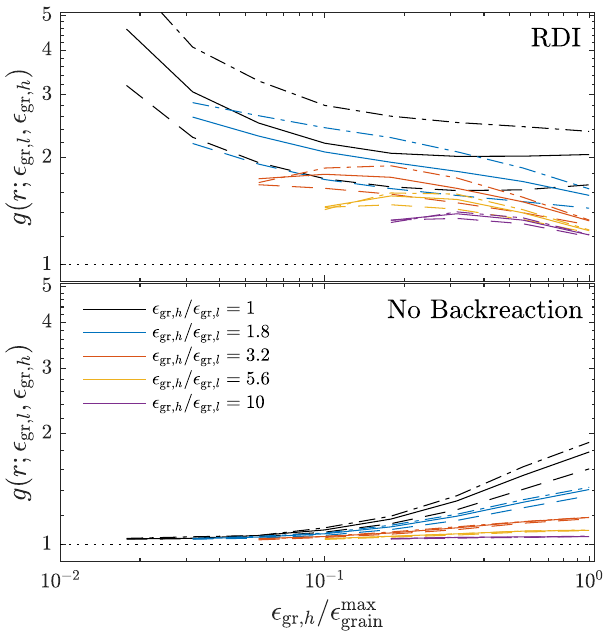}
\caption{Grain-grain density correlation functions $g(r; \abc{l}, \abc{h})$  for the constant-drift 
simulations as a function of the ``primary'' (larger) grain size $\abc{h}$, using the $8$ grain-size bins of \cref{sub: dust binning}. This measures the probability of finding another grain 
of size $\abc{l}$ a distance $r$ from the primary grain of size $\abc{h}$; i.e., it quantifies the mean local density enhancement between grain pairs. Different line colors show different grain-size ratios $\abc{h}/\abc{l}$, as labeled. Different line styles show the correlation at different particle separations $r$; $r=1/256$ (solid lines), $r=1/512$ (dot-dashed lines), and   $r=1/128$ (dashed lines; see text for further information).  We see much 
stronger correlations between grains of all sizes in the RDI simulation, with $g$ rising, rather than falling, with decreasing
grain size for similarly sized grains. The correlations also rise more rapidly 
as $r$ decreases in the RDI case, suggesting that the clumping will be even stronger in the $r\rightarrow 0$ limit relevant for grain-grain collisions.
}
\label{fig: gfunc sda} 
\end{center}
\end{figure}
\begin{figure}
\begin{center}
\includegraphics[width=0.9\columnwidth]{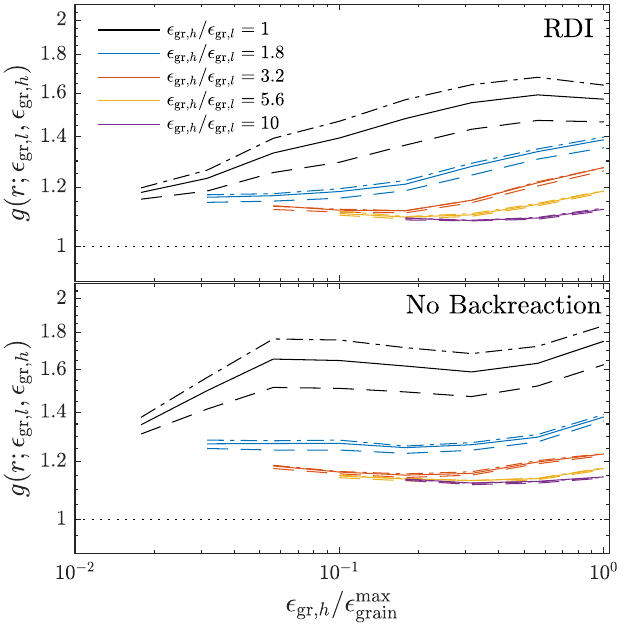}
\caption{As for \cref{fig: gfunc sda}, but for the non-constant-drift simulations. 
The two cases are much more similar than the constant-drift cases, although there
remains a modestly higher correlation between larger grains of different sizes in the RDI turbulence.
}
\label{fig: gfunc ca} 
\end{center}
\end{figure}

\subsubsection{Grain clumping}
The key measure of relative grain clumping is the Radial Distribution Function (RDF), $g(r;\abc{l},\abc{h})$,
which measures the relative probability of finding a grain of size $\abc{l}$ a distance $r$ from a grain 
of size $\abc{h}$. It is normalized such that a spatially homogenous distribution satisfies $g(r;\abc{l},\abc{h})=1$.
If $g(r;\abc{l},\abc{h})>1$ for small $r$, the collision rate of $\abc{l}$ grains with $\abc{h}$ grains will 
be enhanced compared if their distribution were uncorrelated \citep{Maxey1987,Squires1990}. As well as grain collisions, a highly 
clumped grain distribution could have interesting implications  for the  opacity, which  may 
be significantly reduced  compared to a homogenous grain distribution if photons are primarily scattered around the low-density 
regions between dust clumps \citep[see][]{Uli2021RDI}. 

We illustrate the RDF for the constant- and non-constant-drift simulations in \cref{fig: gfunc sda,fig: gfunc ca}, 
comparing the RDI and no-backreaction forced simulations in the top and bottom panels, respectively. We represent $g$
using the same method as \citet{Pan2014b}, setting $\abc{l}<\abc{h}$ and plotting $g(r;\abc{l},\abc{h})$ as a function 
of  $\abc{h}$ for a variety of grain-size  ratios  $\abc{h}/\abc{l}$. Grains are binned before 
computing $g$ according to the method of \cref{sub: dust binning}. The different line styles show 
different $r$, which are computed by including only those grains that lie a distance $r\pm\Delta r$ from 
each other\footnote{The turbulence itself is anisotropic with respect to the drift direction, meaning that $g$ can differ
depending on whether $r$ is in the perpendicular or parallel direction. However, we saw only minor 
differences, when taking this into account, so show only the isotropic version here.}; the solid lines show $r=L/256$  (the equilibrium gas-particle spacing) and $\Delta r =L/1024$, the dashed lines show $r=L/128$ and $\Delta r =L/4096$,
and the dot-dashed lines show $r=L/512$ and $\Delta r =L/1024$
(note that a relatively wider bin in $r$ is needed when $r$ itself is smaller in order to obtain sufficient particle statistics).
Let us first consider the externally forced turbulence constant-drift case, since this is most directly comparable
to previous work, broadly following the expected behavior (c.f.~figure 1 of \citealt{Pan2014b}). The strongest 
turbulent clumping is seen for the largest grains, which is  consistent with the well-documented observation that 
clumping is strongest for particles with ${\rm St}_{\nu}\simeq 1$ \citep{Squires1990,Sundaram1997} and the estimate in \cref{eq: stokes in simulations} that 
all particles have ${\rm St}_{\nu}\lesssim 1$. The maximum of $g\approx 2$ is less than some previous works, and it 
is also clear from the trend that larger grains would clump more strongly. There are a number
of possible causes for this discrepancy: in our driven-turbulence simulations the grains are streaming through the turbulence (unlike previous studies, which used $\aext=0$), which could interfere with the ``sling'' effect that causes turbulent clustering; or the effective viscous scales could be underestimated in the estimates of \cref{subsub: stokes numbers}, thus overestimating ${\rm St}_{\nu}$\footnote{This is supported by examination of the spectra in \cref{fig: spectra}: the velocity spectrum steepens at $k\approx 150/L\approx 24(2\pi/L)$, which is well above the scale of twice the gas particle 
spacing as used in the estimates of \cref{eq: stokes in simulations}.}; or, by not including explicit viscosity, we may not be accurately simulating the sub-viscous-scale flows that determine the clumping of the ${\rm St}_{\nu}\simeq 1$ grains. The decrease in clumping between different-sized particles (compared to particles of the same size) is similar to that shown in \citet{Pan2014b}.

 The  constant-drift RDI  contrasts significantly to the externally forced turbulence. Most obviously, there is much higher clumping for all 
 grain sizes, but particularly for the smallest grains and in the relative correlations between grains of different sizes. This
 strong clumping of small grains is notable given that our simple estimates suggested they have very small Stokes numbers
  (${\rm St}\lesssim 0.01$; c.f. driven case in \cref{fig: gfunc sda}). We also see that our measurements are far from converged, meaning that grains are increasingly clumped at smaller scales, even at scales well below the gas-particle spacing, strongly suggesting that higher-resolution 
  simulations (or reality) would increase $g$ further. Finally, it is worth noting that the general shape of $g(r;\abc{l},\abc{h})$ with $\abc{h}$
  is quite different to those seen in standard hydro turbulence; at least with $\abc{l}\approx\abc{h}$, $g$ becomes independent of $\ad$ for larger $\ad$, rather than decreasing towards $g=1$ at either small or large $\ad$.
  
  The comparison of the non-constant-drift simulations (\cref{fig: gfunc ca}) tells a less interesting story, 
  showing broadly similar grain distributions between the RDI and externally forced turbulence, and less clumping than the 
  constant-drift RDI. This is consistent with only the largest grains driving interesting dynamics in the non-constant-drift RDI, 
  while the smaller grains are primarily passively advected by the flow. The rather low values of $g$ in these cases are likely
  related to the fact that all grains have a slightly 
  different equilibrium streaming velocity $\ws$, meaning that small-scale clumps can be quickly destroyed by grains streaming
  away from each other (this is discussed further below). The convergence to $g\approx1.1$, rather than $g=1$,  is simply because the gas
  is  compressible in these simulations, so that there are always some spatial correlations between grains of different sizes that 
  arise due to their mutual  correlation with the gas density (see \cref{fig:2d hist ca}).



\begin{figure}
\begin{center}
\includegraphics[width=1\columnwidth]{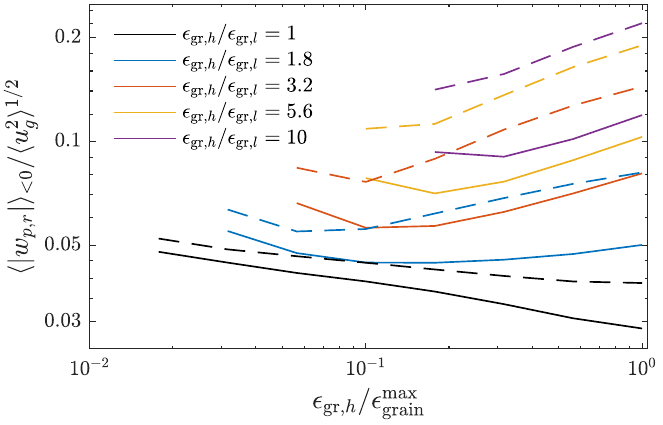}
\caption{Average collision velocity between grain pairs, normalized by the gas velocity dispersion, 
in the constant-drift simulations. Solid lines show RDI turbulence and dashed lines show the externally forced turbulence simulation without backreaction (colors show grain-size ratios, as in \cref{fig: gfunc sda}). We see significantly lower collision velocities in RDI turbulence, particularly between grains of different sizes. Coupled with the larger grain clumping factors in RDI-generated turbulence (\cref{fig:2d hist sda,fig: gfunc sda}), this would significantly enhance 
grain coagulation compared to standard estimates. 
}
\label{fig: wmag sda} 
\end{center}
\end{figure}

\begin{figure}
\begin{center}
\includegraphics[width=1\columnwidth]{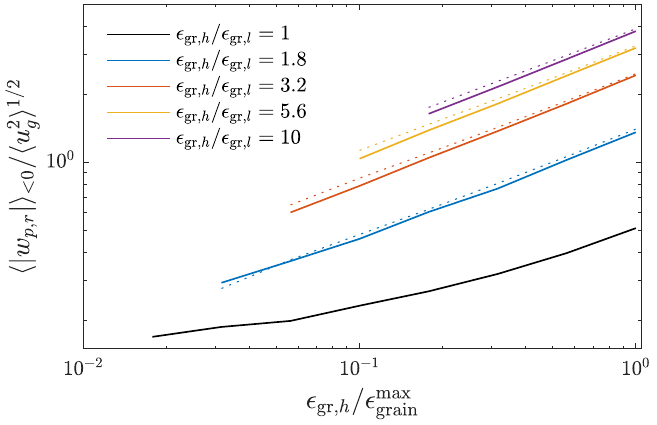}
\caption{Average collision velocity between grain pairs, normalized by the gas velocity dispersion, 
 in the non-constant-drift RDI simulation (solid lines). The dotted lines show the simple prediction \cref{eq: grain collision velocity cda}. In this case, the collision 
velocities are entirely dominated by the different drift velocities of differently sized grains, and the driven and RDI 
cases look very similar. 
}
\label{fig: wmag ca} 
\end{center}
\end{figure}

\subsubsection{Relative velocities}

The second component that is required to estimate the rate and outcome of grain-grain collisions is the relative collision velocity. 
Depending on subtle choices related to the definition 
of the collision kernel, there are a number of relevant measures of collision velocities that can give slightly different results \citep{Wang2000,Pan2014b}. We measure the mean of the value of the 
radial velocity with which particle pairs approach each other, which is needed to compute the ``spherical formulation'' of the collision kernel (unlike, for example, the root-mean-squared relative velocity;  \citealp{Wang2000}). More precisely, for two particles
with velocities $\bm{v}_{1}$ and $\bm{v}_{2}$ and separation vector $\bm{r}$, we define $\bm{w}_{p} = \bm{v}_{2}-\bm{v}_{1}$ and $w_{p,r} = (\bm{v}_{2}-\bm{v}_{1})\cdot\bm{r}/|\bm{r}|$ and take an average $\langle |w_{p,r}|\rangle_{<0}$ over only those pairs that are approaching each other. 
As for the particle RDFs, we plot this separately for different ratios of grain sizes, viz., measure the collision velocities 
of large and small particles, as well as those of similarly sized particles. Collision velocity 
statistics also depend on, and are unconverged in, the particle separation  $r$ (i.e., $\langle |w_{p,r}|\rangle_{<0}=\langle |w_{p,r}(r;\abc{l},\abc{h})|\rangle_{<0}$; \citealp{Pan2014b}). By default we use $r=L/256$
(solid lines in \cref{fig: gfunc sda,fig: gfunc ca}) and comment on this where appropriate. As for the RDFs, 
the collision velocity statistics are not isotropic in either $\bm{w}_{p}$ or $\bm{r}$ (in either the RDI or the no-backreaction cases); however, examination of anisotropy of collisions has not yielded any interesting insights, so we show only the isotropic versions here.

The constant-drift simulations are shown in \cref{fig: wmag sda}. In this case all grains drift at the same average speed, so there is no 
direct contribution to the collision velocity from the grain's equilibrium drift velocity. We 
normalize to gas root-mean-squared velocity in order to make the comparison of the RDI to driven turbulence as apt as 
possible (see \cref{subsub: turb driving level}). Overall, we see a modestly lower collision velocity in the RDI for similar
sized grains, but the difference becomes more significant for collisions between grains of different sizes. 
In other words, collisions between large and small grains are significantly slower on average in RDI turbulence
compared to standard externally forced turbulence. While this might have been anticipated based on our intuitive understanding
that the constant-drift RDI involves gas motions driven by a wide range of dust grains at once, it could 
have interesting implications for the outcome of grain collisions, for instance, by increasing the size to which grains
can grow by sticking  \citep{Blum2018}. Finally, it is worth noting that the collision velocities depend relatively 
strongly on the particle separation $r$, but this dependence is similar for the RDI and externally forced turbulence so we do not consider
it in detail here (see \citealt{Pan2014} for extensive discussion).

The non-constant-drift simulations, which are less interesting, are shown in  \cref{fig: wmag ca}. In these cases, $\ws\sim \ad^{1/2}$ is different for each grain size and dominates over the
turbulent dust velocity dispersion (see \cref{fig:time.evolution.ca}). Grains of different sizes thus collide primarily 
due to their differing drift velocities, with 
\begin{equation}
\bm{w}_{p}=\bm{v}_{2}-\bm{v}_{1}\approx [\ws(\ab{2})-\ws(\ab{1})]\hat{\bm{z}}.\label{eq: grain collision velocity cda}
\end{equation}
For this reason, unrelated to properties of the turbulence, the RDI and externally forced turbulence simulations produce nearly 
identical  $\langle |w_{p,r}|\rangle_{<0}$ and we plot only the RDI case in \cref{fig: wmag ca}. The results  match the estimate \cref{eq: grain collision velocity cda} nearly perfectly (dotted lines; there is an extra 
geometric factor of $1/2$ because only the radial velocity component is computed). While this 
is not surprising, it is nonetheless a potentially relevant physical effect that will significantly enhance the collision 
rate and velocities of grains in outflowing winds in the non-constant-drift regime.

\begin{figure}
\begin{center}
\includegraphics[width=1\columnwidth]{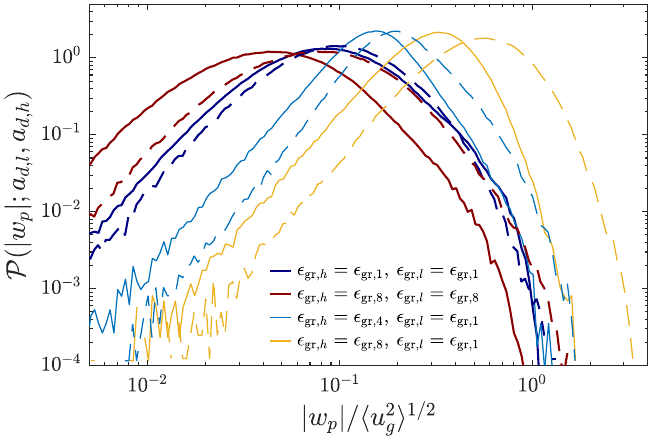}
\caption{PDF of grain collision velocities $|\bm{w}_{p}|$, at particle separation $r=L/256$, for a selection of different grain size pairs 
in the constant-drift simulations. Solid lines show the RDI, dashed lines show the forced turbulence simulation (see
also \cref{fig: wmag sda}).
}
\label{fig: wpdf sda} 
\end{center}
\end{figure}

In addition 
to the mean collision velocities, the PDF of $w_{p}$ is of interest: rare events can have an important impact on the
growth or destruction of grains, for example, by allowing a population to grow beyond particularly important size scales
\citep{Garaud2013,Pan2014a}. We show the PDF of $|\bm{w}_{p}|$, $\mathcal{P}(|\bm{w}_{p}|;\abc{l},\abc{h};r)$ for a variety of grain-size pairs in \cref{fig: wpdf sda}, 
illustrating their   similar shapes in  the RDI and externally forced turbulence for the constant-drift regime.\footnote{The non-constant-drift
PDFs are dominated by the mean drift velocities, so we do not show them here.} Given their clear differences in small-scale 
structure and clumping mechanisms (e.g., \cref{fig: gfunc sda}) this is surprising. One minor difference is a slightly 
steeper high-$w_{p}$ tail (and a slightly flatter low-$w_{p}$ tail) in the RDI (see light-blue and yellow curves), indicating that high-velocity collisions between differently 
sized particles are even less likely than suggested by the collision-velocity average in \cref{fig: wmag sda}. However, 
this seems to be a relatively minor effect.

\subsubsection{The collision rate}


The collision rate between grains of size $\ab{i}$ and $\ab{j}$ is given by  $\nu_{c,ij}=\langle n_{d,i}\rangle\langle n_{d,j}\rangle\Gamma_{ij}$, 
where $\langle n_{d,i}\rangle$ is the number density of species $i$ and $\Gamma_{ij}$ is the collision 
kernel between $i$ and $j$. Using the ``spherical formulation'',\footnote{The alternative ``cylindrical'' formulation gives
very similar results in turbulence simulations, but we have not considered it here \citep{Wang2000,Pan2014b}. } $\Gamma_{ij}=2\pi d^{2} g(d;\ab{i},\ab{j})\langle |w_{p,r}(d;\ab{i},\ab{j})|\rangle_{<0}$, where $d$ is sum of the radii of the grains. This shows 
that $\nu_{c}$ and/or $\Gamma_{ij}$  may be inferred (aside from the $d$ dependence) from \crefrange{fig: gfunc sda}{fig: wmag sda}. We see
that in the non-constant-drift RDI, $\nu_{c}$ is large and strongly dominated by the effect of the
mean drift; this situation will involve a large number of very high velocity collisions between grains of different sizes. In contrast, 
the constant-drift RDI collision rate is  larger than that of externally  forced turbulence  for similarly sized grains (by a factor $\gtrsim 4$ for small grains), and similar
for differently sized grains (since the larger $g$ cancels with the smaller $\langle w_{p,r}\rangle_{<0}$). 
However, the RDI collision velocity is significantly smaller, increasing the probability of slow collisions that lead to grain 
growth as opposed to bouncing, cratering, or fragmentation.

\subsection{Discussion: extensions, limitations, and future work}

{The parameter space of possible  RDIs is very large  \citep{Hopkins2020} and a key limitation 
of our study has been the focus on just two parameter sets of the acoustic RDI for simulation. That said, for the RDI  with 
neutral gas and grains (acoustic RDI turbulence), most of the likely dependencies on other parameters can be inferred from the results here and those of \msh. For larger effective scales (smaller $(2\pi/L)c_{s}t_{s}$), the linear results in \cref{sub: low k linear} show that 
the size distribution of grains becomes unimportant to the instability, suggesting that the non-constant-drift and constant-drift instabilities 
will both behave similarly to the larger scales of the non-constant-drift simulation presented here, or similar cases in \msh. 
For smaller effective scales (larger $(2\pi/L)c_{s}t_{s}$), the constant-drift cases will likely behave similarly but with stronger relative 
clumping for less virulent gas turbulence (\msh), while the non-constant-drift cases will be limited to less virulent quasi-resonant modes involving
only the larger grains (\cref{sub: quasi  modes}). At smaller dust-to-gas-mass ratio. the results of \msh\ suggest that the gas turbulence will become less virulent, but likely cause more {relative} dust clumping (the non-constant-drift RDI will also be less virulent at smaller $\mu$, with the
resonant instability limited to a smaller range of the largest grains; \cref{sub: quasi  modes}). At larger dust-to-gas-mass ratio, 
there is no qualitative change to main features the acoustic RDI (as occurs for the streaming instability; \citealp{Youdin2005,Squire2020a});
rather, the larger mass of the dust simply drives stronger gas turbulence (\msh). 
Finally, it is also worth mentioning that while GIZMO seems to be able to capture the 
linear growth rates of the polydisperse acoustic RDI relatively accurately (see, e.g., \cref{fig: nl lin comparison k theta}), 
exploring the detailed convergence to linear predictions in different regimes with different grain-size distributions is a complex task 
beyond the scope of this work (see, e.g., \citealt{Paardekooper2021,Zhu2021}). While unlikely to affect our results here, given the
dominance of the large-scale modes in the non-constant-drift simulation, there may be important effects at smaller scales and/or smaller $\mu$,
and a more detailed study of numerical convergence and/or comparison to other codes would be important for exploring such cases.  }

{We have also not studied in detail the (likely common)
situation where the spectrum of dust grains covers both the constant- and non-constant-drift regimes. Based on linear calculations (see \cref{app: linear}), it is reasonable to 
surmise  that this will behave like the constant-drift RDI for the relevant range of grains (albeit with 
a somewhat reduced dust-to-gas-mass ratio).  }

{A rather technical source of uncertainty in our results relating to dust-dust collisions concerns the approach to the
zero-particle-separation limit. As discussed extensively in \citet{Pan2013,Pan2014b,Haugen2021}, a quantitative measurement
of the collision kernel requires a careful convergence study in particle separation, which is not achieved here (e.g., compare the different line styles in 
\cref{fig: gfunc sda}). This must take into 
account both  explicit dissipative effects, which damp gas motions at small scales, and the different physical contributions 
to relative grain-grain velocities.\footnote{\revchng{\citet{Wilkinson2006} introduced the decomposition of the grain-grain collision rate into 
its continuous and ``caustic'' parts, where the caustic part accounts for  collisions that result from grains being slung out from neighbouring 
eddies.  In order to separate these contributions, which is needed to estimate the collision rate in the zero-particle-separation limit, one needs
a reasonable model for the behavior of a grain in a representative turbulent eddy \citep{Pan2013,Pumir2016}. }}. This is well beyond the scope of this work and
likely a  difficult task for realistic  application to astrophysical objects with RDI-generated turbulence. In particular,
as well as a better theoretical framework with which to understand the strong clumping even at small Stokes
numbers, the acoustic RDI effectively forces the turbulence down to the viscous scales, changing in character with scale. These issues
make it difficult to apply Stokes-number-based self-similarity  arguments, as usually applied to understand passive grains in turbulence. Thus, the result that  constant-drift RDI turbulence should 
be particularly a effective nursery for grain growth is qualitative at this stage, and we refrain from making quantitative 
estimates for the collision kernel. }

{That said, the most  relevant and important uncertainty of our study is our
neglect of grain charge and gas magnetization. The neutral cases explored here provide 
a reasonable approximation for cooler, denser regions, for instance around  AGB stars, or in some parts of the cool ISM;
but, in most astrophysical scenarios where RDIs apply, grain charge and gas magnetization are expected to play a key role. 
Further, magetized-RDI turbulence  causes significantly stronger clumping than acoustic-RDI turbulence, even 
in the regime where drag forces are stronger than Lorentz forces and/or in the non-constant-drift 
regime.\footnote{It remains unclear exactly why this is, although 
there are many more resonant instabilities available in the magnetic case due to the wider variety of waves 
and the dust's gyromotion.} A selection of cases is presented in \cite{Hopkins2021}, mostly focusing
on  the more complex 
situation of a stratified wind driven from the base: extremely strong clumping is seen even in non-constant-drift 
cases (their ``\textbf{-Q}'' simulations; see e.g., their figure 9) in stark contrast to our results here,  a difference that is related to the magnetization and not the
stratification of the system (see e.g., their figure 13; this is also seen in unstratified magnetized simulations).
From these results it is clear that magnetization will be a key parameter for RDI-turbulence induced grain clumping;
however, given the complexity of these cases and the yet wider parameter space to explore, we leave such studies to future work. }

{Finally, it is worth mentioning that by neglecting explicit stratification of the system, we are also 
neglecting other possible gas and/or radiative instabilities that can occur in dust-driven winds for some systems \citep[see, e.g.,][]{Woitke2006,Krumholz2012}.
Generically, such instabilities operate on much larger scales than the acoustic RDIs considered here, although it is plausible 
that they could create turbulence that influences RDI development in some circumstances. 
Again, such issues are addressed more explicitly in the stratified simulations of \cite{Hopkins2021}.}

\section{Conclusions}

This paper has presented an in-depth study of the   ``acoustic Resonant Drag Instability''  (\citealp{Squire2018}; \hs), which 
is driven by the interaction of an outflowing population of dust grains with compressible gas motions. 
The acoustic RDI is expected to operate in a variety of astrophysical scenarios, for example,
in the presence of a radiation source that couples more strongly 
to dust grains than to the gas. This accelerates grains outwards, often to supersonic velocities, 
which (in addition to driving a gas outflow) destabilizes the RDI  (\hs).   
In cooler, denser gas, such as that in molecular clouds, AGB-star winds, or around AGN, the grain charge and MHD effects
are not necessarily dominant, and the system may be well approximated by considering a neutral gas and neutral grains (the ``acoustic'' RDI).
The novel feature of this work has been the inclusion of a wide spectrum of grain sizes -- a factor of $100$ in grain 
radius $\ad$ in our numerical simulations -- which has not been included in previous
studies but is clearly an important feature of realistic  systems \citep{Draine2010}.
In our numerical study, rather than surveying a wide parameter space of different simulations, we 
have focused in detail on two representative cases that can apply adequately well to a variety of astrophysical processes. By 
comparing directly to simulations of externally forced turbulence, this allowed
us to consider in more detail aspects of dust and gas structure in the RDI, and how these might 
influence important processes such as grain growth.

With a spectrum of grains, there are two qualitatively different regimes of the RDI depending on 
how the grain's acceleration (imparted, for example, by an external radiation field) scales with 
the grain radius. In the \emph{constant-drift} regime, applicable for grains larger than the wavelength 
of a radiation field ($\ad\gtrsim \lambda_{\rm rad}$), all grains drift through the gas with the 
same velocity in the (quasi-)equilibrium. In the  \emph{non-constant-drift} regime, applicable for small 
grains in long-wavelength radiation fields ($\ad\lesssim \lambda_{\rm rad}$), or when 
the acceleration difference between the dust and gas arises from an external force on the gas, the grain drift velocity increases with grain size, stretching across  a wide range of values in realistic scenarios. 

We show in \cref{app: linear}, which presents analytic and numerical calculations of the {linear} RDI growth rate 
with a spectrum of grains, that these different regimes strongly 
influence the behavior of the acoustic RDI. Generally, the RDI is more virulent, with faster growth rates and behavior 
that is very similar to the single-grain-size case, in the constant-drift regime. Our simulations
show that this linear behavior carries over into the nonlinearly turbulent regime also:
 the non-constant-drift RDI, although strongly unstable with a saturated state that shares key features
 of the linear instability, develops into turbulence without strong
 correlations between grains of different size (like externally forced turbulence); 
the constant-drift RDI is very different, with much stronger spatial correlations between 
small grains and those of different sizes, along with slower grain-grain collision velocities.   These
differences, which imply a high rate of low-velocity collisions in the constant-drift 
RDI, suggest that  constant-dust-drift  outflows could be highly effective sites
for dust  growth  through collisions, while the opposite is likely true in the non-constant-drift regime because 
grain-grain collisions are dominated  by the (fast) mean drift between grains of different sizes. 
{Another interesting conclusion in the constant-drift regime
is the  strong clumping of small grains, even though their turbulent Stokes numbers remain well below one (see \cref{subsub: stokes numbers}). 
This highlights the fact
 that the clumping mechanism in RDI-generated turbulence is quite different -- and much more efficient for similar velocity dispersions --  to standard (Kolmogorov) turbulence \citep{Pumir2016},
 even though the turbulent velocity spectra are relatively similar (\cref{fig: spectra}).  }

\acknowledgements{We thank Eric Moseley and Darryl Seligman for helpful discussion. 
Support for JS  was
provided by Rutherford Discovery Fellowship RDF-U001804 and Marsden Fund grant UOO1727, which are managed through the Royal Society Te Ap\=arangi. Support for JS, PFH, and SM was provided by NSF Collaborative Research Grants 1715847 \& 1911233, NSF CAREER grant 1455342, and NASA grants 80NSSC18K0562 and JPL 1589742. Numerical simulations were
run on the Caltech compute cluster ``Wheeler,''  and with  allocation TG-AST130039 from the 
 Extreme Science and Engineering Discovery Environment (XSEDE), which is supported by National Science Foundation grant number ACI-1548562.}

\paragraph*{Data availability}{The simulation data presented in  this article is available on  request to JS. A public version of the GIZMO code is available at \href{http://www.tapir.caltech.edu/~phopkins/Site/GIZMO.html}{http://www.tapir.caltech.edu/~phopkins/Site/GIZMO.html}.}





\appendix

\label{ --  --  --  --  --  APPENDIX  --  --  --  -- --}

\section{Linear behaviour}\label{app: linear}

%

In this extended appendix, we analyse key properties of the acoustic resonant drag instability 
in the presence of a spectrum of grain sizes. We will show that its behaviour depends
strongly on the drift regime described in \cref{sub: acceleration} and \cref{tab: regime}: in the 
constant-drift regime (grain-size dependent acceleration) the instability 
is very similar to the single-grain acoustic RDI  (\hs) because all grains share the 
same drift velocity; in the 
non-constant-drift regime (acceleration independent of grain size), the instability is quite different
 and generally less virulent at small scales. 
 
{ Although we consider only the acoustic RDI here, it is worth noting that a number 
 of the features we discuss are shared by other RDIs, which can exhibit  a diversity of different 
 behaviors. 
A spectrum of grains has been shown to dramatically affect the linear growth of the $\mu<1$
streaming instability in protoplanetary disks \citep{Youdin2005}, which 
is also an RDI. Specifically, for smaller grain sizes, \citep{Krapp2019} found that growth 
rates were a strong function of the number of species $N_{d}$ used to discretize the grain distribution,
in some cases being unconverged even for $N_{d}>2048$. Subsequent works \citep{Paardekooper2020,Zhu2021,McNally2021}
have explored this further, showing that the instability is robust only for large stopping times or $\mu\gtrsim1$, 
also depending somewhat on the size distribution of grains.
Similar non-convergent behavior is
seen for some mode angles of the acoustic RDI discussed below.
However, such behavior is not by any means generic to all RDIs. \cite{Krapp2020} showed that the
``settling instability'' does not suffer from the same convergence issues, 
despite its similarity to the streaming instability (the only difference being the direction of dust drift;  \citealp{Squire2018a}).
Similarly,  \cite{Squire2021} computed analytically the  growth rates of magnetized gyroresonant RDIs with charged
dust grains, which converge rapidly and easily with $N_{d}$ to fast growth in the continuum limit. 
This wide variety of behaviors is also exhibited by our calculations below, with some ranges of 
mode angles and wavenumbers converging very slowly in $N_{d}$ and some not. More general 
exploration of these issues is left to future work.}

  We first describe the basic set up of the calculations
  in \cref{app: linear general}, then specialise to the simpler constant-drift case in \cref{app: grain dep accel}, before considering the more general non-constant-drift regime in \cref{app: grain indep accel}. While realistic scenarios
  will often involve a population of grains that stretches across both regimes, we do not consider this
possibility here for simplicity. We also do not consider the magnetic RDIs studied in \citet{Hopkins2018a} due to 
their complexity and the wide range of parameter regimes.

\subsection{General considerations}\label{app: linear general}
Our theoretical approach will involve computing linear growth rates by starting from a finite number ($N_{d}$) of grain-size species. We then 
take the limit as $N_{d}\rightarrow \infty$ with the total mass of grains fixed to obtain results valid in the continuous limit. {This is 
by no means the most efficient approach in all cases \citep[see][]{Paardekooper2021}, but is physically enlightening for understanding the physical 
causes of different behaviors.}
We thus define the properties of discrete grain species $j$: the dust-to-gas-mass ratio $\mu_{j}$ (with $\mu_{0}=\sum_{j}\mu_{j}$), the 
relative drift velocity $\wsj=|{\bf w}_{s,j}|/c_{s}$, and the stopping time $\tsj$, which become functions of grain size $\ad$ in the continuous limit. We assume
that the quasi-equilibrium involves pressureless  dust (i.e., no dust velocity dispersion for all grain sizes), 
\begin{equation}
\int_{\ad}^{\ad+d\ad} d\ad\,d\bm{v}\,[\bm{v}-\bm{v}_{d}(\ad)] [\bm{v}-\bm{v}_{d}(\ad)] f_{d}(\ad;\,\bm{x},\bm{v})=0,\label{eq: pressureless dust}
\end{equation}
which implies that the the dust evolution equation \eqref{eq: f dt} is equivalent to fluid equations for its first two moments: 
\begin{gather}
\ddt{\rho_{d}} + \nabla\cdot(\rho_{d}\bm{v}_{d})=0,\label{eq: dust fluid rho}\\
\ddt{\bm{v}_{d}} + \bm{v}_{d}\cdot\nabla(\bm{v}_{d}) = -\frac{\bm{v}_{d}-\bm{u}_{g}}{t_{s}(\ad,\bm{v}_{d})}.\label{eq: dust fluid v}
\end{gather}
Here $\rho_{d}$ and $\bm{v}_{d}$ are each functions of $\ad$, so $N_{d}$ copies of \cref{eq: dust fluid rho,eq: dust fluid v} are needed, each with different stopping times $\tsj=t_{s}(a_{d,j})$ and $\wsj$.
We then linearise \cref{eq: dt cont,eq: dt mom,eq: dust fluid rho,eq: dust fluid v} about the quasi-equilibrium in which the dust and gas 
accelerate linearly together at the same rate, but allowing for species-dependent velocity offset  (see \hs\ for more information). Denoting the quasi-equilibrium quantities with $\langle \cdot\rangle$, 
and moving to the frame in which the gas is stationary this leads to a set of $3(N_{d}+1)$ linear equations
for the gas density perturbation $\delta \rho_{g} = (\rho_{g}-\langle \rho_{g}\rangle)/\langle \rho_{g}\rangle$, gas velocity perturbation
$\delta \bm{u}_{g}=\bm{u}_{g} $, dust density perturbation $\delta \rho_{d,j}=(\rho_{d,j}-\langle \rho_{d,j}\rangle)/\langle \rho_{d,j}\rangle$, and dust velocity perturbation $\delta \bm{v}_{d,j}=\bm{v}_{d,j}-\langle \bm{v}_{d,j}\rangle = \bm{v}_{d,j} - 
\wsj c_{s}\hat{\bm{z}}$:
\begin{gather}
\frac{\partial}{\partial t}\begin{pmatrix}
\delta \bm{D}_{1} \\ \vdots \\ \delta\bm{D}_{j}  \\ \vdots \\ \delta \bm{F}
\end{pmatrix}
= 
\mathbb{T}\cdot 
\begin{pmatrix}
\delta\bm{D}_{1} \\ \vdots \\ \delta \bm{D}_{j}  \\ \vdots \\ \delta\bm{F}
\end{pmatrix}
=
\begin{pmatrix}
\mathcal{A}_{1}  & 0 & 0 & 0 &  \mathcal{C}_{1} \\
0   & \ddots & 0 & 0 & \vdots \\
0 & 0 & \mathcal{A}_{j} & 0 & \mathcal{C}_{j}\\
0   & 0 & 0 & \ddots &   \vdots\\
\mathcal{T}_{1}   & \dots & \mathcal{T}_{j}  & \dots &   \mathcal{F}\\
\end{pmatrix}\cdot 
\begin{pmatrix}
\delta\bm{D}_{1} \\ \vdots \\ \delta \bm{D}_{j}  \\ \vdots \\ \delta\bm{F}
\end{pmatrix},\nonumber \\
\delta \bm{D}_{j} = \begin{pmatrix}
 \delta \rho_{d,j}\\ \delta {v}_{x,j} \\  \delta {v}_{z,j} 
\end{pmatrix},\quad 
\delta \bm{F} = \begin{pmatrix}
\delta \rho_{g} \\ \delta u_{x} \\ \delta u_{z} 
\end{pmatrix},\nonumber\\
\mathcal{A}_{j} = \begin{pmatrix}
-i k_z w_{s,j} & -i k_x & -i k_z  \\
  0 & -i k_z w_{s,j}-\frac{1}{t_{s,j}} & 0   \\
 0 & 0 & -i k_z w_{s,j} -\frac{\tilde{\zeta}_{w,j}}{t_{s,j}} \\
\end{pmatrix},\nonumber \\
\mathcal{C}_{j} = \begin{pmatrix}
 0 & 0 & 0 \\
 0 & \frac{1}{t_{s,j}} & 0 \\
 -\frac{\zeta _{s,j} w_{s,j}}{t_{s,j}} & 0 & \frac{\tilde{\zeta }_{w,j}}{t_{s,j}} 
\end{pmatrix},\quad
\mathcal{T}_{j} = \mu_{j}\begin{pmatrix}
0 & 0 & 0  \\
  0 & \frac{1}{t_{s,j}} & 0   \\
 \frac{w_{s,j}}{t_{s,j}} & 0   & \frac{ \tilde{\zeta }_{w,j}}{t_{s,j}}
\end{pmatrix},\nonumber \\
\mathcal{F} = \begin{pmatrix}
 0 & -i k_x & -i k_z \\
 -i c_{s}^{2}k_x & -\sum_{i}\frac{\mu _i}{t_{s,i}} & 0 \\
-i c_{s}^{2}k_z +\sum_{i}\frac{\mu _i \left(\zeta _{s,i}-1\right)
   w_{s,i}}{t_{s,i}} & 0 & -\sum_{i}\frac{\mu _i \tilde{\zeta}_{w,i}}{t_{s,i}}
\end{pmatrix}
\label{eq: full matrix}
\end{gather}
Here we have considered, without loss of generality, ${\bf w}_{s,j}\propto \hat{\bm{z}}$ and two-dimensional spatial perturbations with wavenumber 
$\bm{k}=(k_{x},0,k_{z})$, implying the spatial form of any perturbation is $\propto \exp(ik_{x}x+ik_{z}z)$. For convenience 
below, we will use $k$ and $\theta_{k}$ as variables, with $k_{x}=k\sin\theta_{k}$ and $k_{z}=k\cos\theta_{k}$. The parameters
$\zeta_{s,j}$ and $\zeta_{w,j}$  are the 
 density and velocity dependencies of the stopping 
time, respectively, which are defined from 
\begin{equation}
\frac{\delta \tsj}{\langle \tsj\rangle} = - \zeta_{s,j}\frac{\delta \rho_{g}}{\langle \rho_{g}\rangle}-\zeta_{w,j}\frac{{\bf w}_{s,j}\cdot(\delta \bm{v}-\delta \bm{u}_{g})}{c_{s}^{2}\wsj^{2}},\label{eq: zeta defs}
\end{equation}
and $\tilde{\zeta}_{w,j}\equiv {\zeta}_{w,j}+1$. For Epstein drag,
\begin{equation}
\zeta_{s,j}=\frac{\gamma+1+2a_{\gamma} \wsj^{2}}{2+2a_{\gamma} \wsj^{2}},\quad \zeta_{w,j} = \frac{a_{\gamma} \wsj^{2}}{1+a_{\gamma} \wsj^{2}},\label{eq: zeta def}
\end{equation}
with $a_{\gamma}=9\pi\gamma/128$.
The structure of the matrix \eqref{eq: full matrix}, which is nonzero only in the blocks shown, encapsulates the physics
of the multi-species dust and gas system;  a given species of dust is governed by its own dynamics (block $\mathcal{A}_{j}$), the influence of the gas on species $j$ (block $\mathcal{C}_{j}$), the influence of species $j$ on the gas  (block $\mu_{j}\mathcal{T}_{j}$), and the dynamics of the gas itself  (block $\mathcal{F}$). 
Importantly, there is no direct coupling between grain species, which would manifest in the matrix structure as off-diagonal blocks other than $\mathcal{C}_{j}$ and $\mathcal{T}_{j}$. The system exhibits linear instability if  eigenvalue(s) of the matrix $i\,\mathbb{T}$ have a positive imaginary part.

To simplify some integrals in the sections below, we will assume grains with an MRN mass distribution ($d\mu/d\ln \ad\propto \ad^{1/2}$ between $\amin$ and $\amax$), 
which implies  
\begin{equation}
\frac{d\mu}{d\ad} = \frac{\mu_{0}}{2}\frac{\ad^{-1/2}}{(\amax)^{1/2}-(\amin)^{1/2}},
\end{equation}
where $\mu_{0}$ is the total dust-to-gas mass ratio. We also assume dust is drifting in the supersonic regime with Epstein drag, which implies
 $t_{s}\propto \ad$ and $\ws\sim \mathrm{const.}$ for the constant-drift regime, and 
$t_{s}\propto \ad^{1/2}$, $\ws\propto \ad^{1/2}$ for the non-constant-drift regime.

\subsection{Constant-drift regime}\label{app: grain dep accel}

In this section we specialise  to the case where the acceleration imparted on grains by the radiative forcing scales as $\aext\propto 1/\ad$, as
relevant when $\ad\lesssim \lambda_{{\rm rad}}$ (see \cref{sub: acceleration}; \citealp{Weingartner2001b}). Importantly, in this regime,
all grains share the same drift velocity $w_{s,j}$, which significantly simplifies the linear mode structure compared to the more general case. The growth 
rate of the fastest-growing modes at moderate $k$  can be straightforwardly obtained using arguments similar to those of \citet{Squire2018}, which are 
based on perturbing the eigenvalues of the matrix in the small parameter $\mu$. To generalize to the case with a spectrum
of grains, we assume that all $\mu_{j}$ are of the same order such that $\mu_{j}=\mu \bar{\mu}_{j}$ with $\bar{\mu}_{j}\sim\mathcal{O}(1)$, 
then treat the parts of $\mathbb{T}$ that contain $\mu$ ($\mathcal{T}_{j}$ and part of $\mathcal{F}$) as a perturbation; i.e., $\mathbb{T}=\mathbb{T}_{0}+\mu\mathbb{T}^{(1)}$. One then 
notes that at the specific ``resonant'' mode angle, when $w_{s,j}\cos\theta_{k}=1$, 
the unperturbed system ($\mathbb{T}|_{\mu=0}$) is $N_{d}+1$ degenerate and 2-fold defective (there are only $N_{d}$ eigenvectors). Physically, this is 
where the sound-wave frequency $\omega_{F}=k\, c_{s}$ matches the streaming frequency ($\omega_{D,j}=k_{z}\, c_{s}\wsj$) of all dust species. By standard linear algebra 
arguments \citep{Kato2013,Moro2002}, in order to compute the eigenvalues of the perturbed system, one should perform a similarity transformation of $\mathbb{T}_{0}$ and $\mathbb{T}^{(1)}$ to form 
a block that is as close to diagonal as possible with $\omega_{0}=\omega_{F}=\omega_{D,j}$ along the diagonal. This  yields a transformed  $\mathbb{T}_{0}$ of the form 
\begin{equation}
\tilde{\mathbb{T}}_{0} = 
\begin{pmatrix}
\omega_{0}  & 0 & 0 & 0 & c_{1} \\
0   & \ddots & 0 & 0 & \vdots \\
0 & 0 & \omega_{0} & 0 & c_{j}\\
0   & 0 & 0 & \ddots &   \vdots\\
0   & \dots & 0 & \dots &   \omega_{0}\\
\end{pmatrix},\label{eq: T0 matrix transformation }
\end{equation}
while $\mathbb{T}^{(1)}$ transforms to
\begin{equation}
\tilde{\mathbb{T}}^{(1)}=\begin{pmatrix}
\cdot & \cdot & \cdot & \cdot & \cdot \\
\cdot   & \cdot & \cdot & \cdot & \cdot \\
\cdot & \cdot& \cdot & \cdot & \cdot\\
\cdot   & \cdot & \cdot & \cdot &   \cdot\\
t_{F,1}   & \dots & t_{F,j}  & \dots &  t_{F,F}\\
\end{pmatrix}\label{eq: T1 matrix transformation }
\end{equation}
(where it transpires that most of $\tilde{\mathbb{T}}^{(1)}$ will not be needed).
Here $c_{j}=\xi_{D,j}^{L}\cdot\mathcal{C}_{j}\cdot\xi_{F}^{R}$ and  $t_{F,j}=\xi_{F}^{L}\cdot\mathcal{T}_{j}\cdot\xi_{D,j}^{R}$ are now scalars, with 
$\xi_{F}^{R}$ ($\xi_{F}^{L}$) the right (left) eigenvector of $\mathcal{F}|_{\mu=0}$ and $\xi_{D,j}^{R}$ ($\xi_{D,j}^{L}$) the right (left) eigenvector of $\mathcal{A}_{j}$. One then computes the eigenvalues of $\tilde{\mathbb{T}}_{0}+\mu\tilde{\mathbb{T}}^{(1)}$ by noting that for $\mu\ll1$, the dominant balance solution for $\omega^{(1)}=\omega-\omega_{0}$ scales as $\sim\mu^{1/2}$ and is
\begin{equation}
\omega^{(1)}=\pm \mu^{1/2}\left( \sum_{j} t_{F,j} c_{j}\right)^{1/2} + \mathcal{O}(\mu).
\end{equation}
Thus, while the full $\mathcal{O}(\mu)$ solution for $\omega^{(1)}$ is very complex and depends on all components of $\tilde{\mathbb{T}}^{(1)}$ ($t_{F,F}$ and other components in \cref{eq: T1 matrix transformation }), the growth rate at resonance ($\wsj\cos\theta_{k}=1$) is straightforward to 
derive and  larger than the growth rate at any other $\theta_{k}$ (because $\mu^{1/2}\gg \mu$ for $\mu\ll1$). Inserting 
the matrices from \cref{eq: full matrix} to compute $t_{F,j}$ and $c_{j}$, 
we find a remarkably simple expression for the  growth rate at the resonant mode angle $\cos\theta_{k}=1/\wsj$,
\begin{align}
\Im(\omega)&\approx \frac{1}{2}\left[k\,c_{s} \sum_{j} \mu_{j}\frac{1}{\tsj}\left(1-\frac{\zeta_{s,j}}{\tilde{\zeta}_{w,j}}\right)\right]^{1/2}\nonumber \\
& \approx \frac{1}{2} \left[ k\,c_{s} \int d\ad\, \frac{d\mu}{d\ad}\frac{1}{t_{s}(\ad)} \left(1-\frac{\zeta_{s}}{\tilde{\zeta}_{w}}\right)\right]^{1/2}\nonumber \\
& \approx \frac{\mu_{0}^{1/2}}{2} \left[ \frac{k\,c_{s}}{t_{s,{\rm max}}} \sqrt{\frac{\amax}{\amin}} \left(1-\frac{\zeta_{s}}{\tilde{\zeta}_{w}}\right)\right]^{1/2},\label{eq: size dep final ans}
\end{align}
where we convert to a continuous distribution on the second line, and specialise to an MRN distribution of supersonically drifting grains on the third line
(with $t_{s,{\rm max}}$  the stopping time of the maximum  grain size). Note
that, like $w_{s}$, $\zeta_{s}$ and $\zeta_{w}$ are independent of $\ad$ in the constant-drift regime.

As in the single-grain case (see \hs), \cref{eq: size dep final ans} is valid  for $\mu\ll k c_{s}t_{s}(\ad)\ll \mu^{-1}$. At very short wavelengths ($k\,c_{s}t_{s}\gg\mu^{-1}$) the growth rate 
keeps a similar structure but transitions to scaling as $\sim \mu^{1/3}k^{1/3}$; we do not consider this in detail because
it is a difficult regime to study numerically (see \hs\ and \msh). At very long wavelengths ($k\,c_{s}t_{s}\ll\mu$) the growth 
rate loses its resonant character and scales with $\sim \mu^{1/3} k^{2/3}$; this case is considered below  (\cref{sub: low k linear}) because it 
the growth rate does not depend strongly on all grains having the same $w_{s}$ (see \cref{eq: low k growth rate cont size dep}).

\begin{figure}
\begin{center}
\includegraphics[width=1.0\columnwidth]{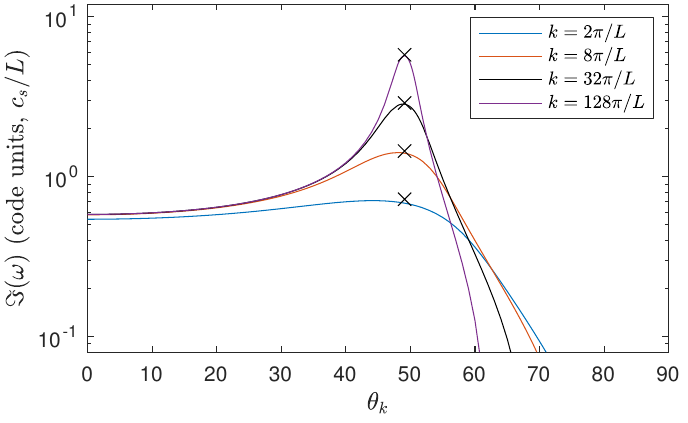}
\caption{Linear growth rate as a function of mode angle $\theta_{k}$ for the parameters of the constant-drift simulation from 
the main text. Different 
coloured curves show wavenumbers relevant to the largest scales of the box (blue curve), down to several resolution elements (purple curves).
Black crosses show the analytic prediction \eqref{eq: size dep final ans}, illustrating the accurate prediction of the
fastest growing mode angle and growth rate. The numerical growth-rate calculation is discretised with $N_{d}=128$ species of grains, but results are almost independent
of $N_{d}$.}
\label{fig:linear adep}
\end{center}
\end{figure}

We plot linear growth rates for the same parameters as the nonlinear constant-drift RDI simulation in the main text as a function of $\theta_{k}$ in \cref{fig:linear adep}. The different lines show  a number of 
different $k$ relevant to the simulation size. The 
analytic result, \cref{eq: size dep final ans}, is shown with the black crosses, and clearly provides a very accurate approximation to the 
maximum growth rate and wave number.  The general features of the instability are very similar to the single-grain case, with a resonant
peak that scales as  $\Im(\omega)\sim\mu_{0}^{1/2}$ and
 increases in sharpness  towards small scales.

\subsection{Non-constant-drift regime}\label{app: grain indep accel}

In the non-constant-drift regime, which is appropriate when $\ad\gtrsim\lambda_{\rm rad}$, or when a force 
accelerates the gas rather than the dust,
the population of grains no longer share the same drift velocity. 
This significantly complicates the analysis of the resonant modes compared to that of \cref{app: grain dep accel}, and useful analytic expressions
for the growth rates are much harder to obtain. Further, in some regimes the growth rate scales with the number of
dust species $N_{d}$, and care is required even for the interpretation of numerically computed growth 
rates (this also occurs in many regimes for the streaming instability in protoplanetary disks, see \citealt{Krapp2019,Paardekooper2021}). For this reason, we provide only a cursory survey of different modes and regimes here, pointing out some
key features of how the growth rate varies with wavelength, but forgoing a full detailed analysis.  As in \hs\
we will organise the results by the type of mode, describing the most important modes across different regimes. Further details of the mode structure, growth rates, and possible instabilities that can occur with different drag laws are given in \hs.

\subsubsection{Low-$k$ modes}\label{sub: low k linear}
At very long wavelengths, $k\,c_{s}t_{s}\ll\mu$, the system exhibits a non-resonant instability that is insensitive to the distribution of grain velocities and sizes. One can derive its growth rate using the
 same analysis as \hs, starting from the
 dispersion relation (characteristic polynomial of $\mathbb{T}$).  Making the substitution $\kappa_{\|}=k\,c_{s} \cos\theta_{k}$ and assuming $w_{s,i}\sim \mathcal{O}(1)$, one assumes $\omega\sim \kappa_{\|}^{2/3}$ and expands the dispersion relation in $\kappa_{\|}\ll1$. This process,
 which is straightforwardly carried out for $N_{d}=1$, $2$, $3$, or $4$ using a computer algebra system, shows that the 
 growth rate for arbitrary $N_{d}$ is the obvious generalisation of the single-species case,
\begin{equation}
\Im(\omega)\approx\frac{1}{2}\left[ \frac{1}{1+\sum_{j}\mu_{j}}\sum_{j}\mu_{j}\frac{\wsj^{2}}{\tsj}\left(1-\frac{\zeta_{s,j}}{\tilde{\zeta}_{w,j}}\right)( k\,c_{s} \cos\theta_{k})^{2}\right]^{1/3}\label{eq: low-k growth rate}
\end{equation}
(taking $\tilde{\zeta}_{w,j}>\zeta_{s,j}$ as relevant to Epstein drag).
Replacing the sum with an integral and assuming $\mu\ll1$ then yields the growth rate for a continuous spectrum of grains,
rate
\begin{equation}
\Im(\omega) \approx\frac{1}{2}\left[ \int d\ad\,\frac{d\mu}{d\ad} \frac{w_{s}^{2}(\ad)}{t_{s}(\ad)}\left( 1-\frac{\zeta_{s}(\ad)}{\tilde{\zeta}_{w}(\ad)} \right)( k\,c_{s} \cos\theta_{k})^{2} \right]^{1/3}.\label{eq: low k growth rate cont}
\end{equation}
Considering first the non-constant-drift regime, \cref{eq: low k growth rate cont} can be simplified to 
\begin{equation}
\Im(\omega) \approx \frac{\mu_{0}^{1/3}}{2}\left[ \frac{w_{s,{\rm max}}^{2}-w_{s,{\rm min}}^{2}}{t_{s,{\rm max}}-t_{s,{\rm min}}}\left( 1-\frac{{\zeta_{s}}}{{\tilde{\zeta}_{w}}} \right)( k\,c_{s} \cos\theta_{k})^{2}\right]^{1/3},\label{eq: low k growth rate cont size indep}
\end{equation}
by assuming constant-$\aext$ supersonic drift with an MRN distribution. (To obtain a simple result for the integral, we neglect the grain-size dependence of $\zeta_{s}$ and $\zeta_{w}$; the dependence is  minor for supersonic drift, with $\zeta_{s}/\tilde{\zeta}_{w} =1/2+\mathcal{O}(w_{s}^{-2})$). 

{Because} the growth mechanism is non-resonant, the same argument  also applies in the
constant-drift regime when $k\,c_{s}t_{s}\ll\mu$, in which case \cref{eq: low k growth rate cont} simplifies to 
\begin{equation}
\Im(\omega) \approx \frac{\mu_{0}^{1/3}}{2}\left[ \frac{w_{s}^{2}}{t_{s,{\rm max}}}\sqrt{\frac{\amax}{\amin}}\left( 1-\frac{{\zeta_{s}}}{{\tilde{\zeta}_{w}}} \right)( k\,c_{s} \cos\theta_{k})^{2}\right]^{1/3}.\label{eq: low k growth rate cont size dep}
\end{equation}
\Cref{eq: low k growth rate cont size indep,eq: low k growth rate cont size dep} match numerical solutions of the dispersion relation perfectly (not shown).

\begin{figure}
\begin{center}
\includegraphics[width=1.0\columnwidth]{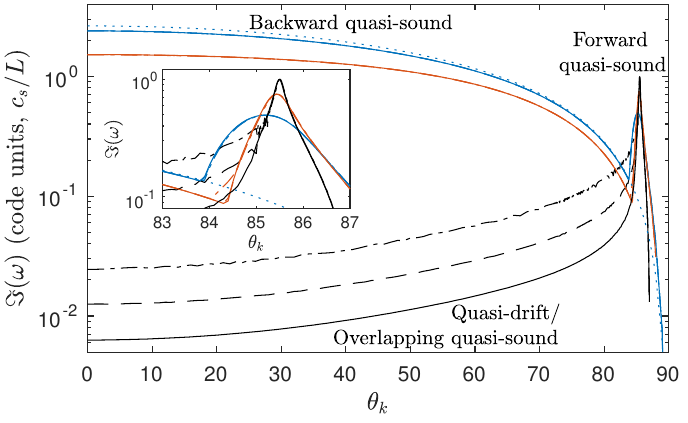}
\caption{Linear growth rate as a function of mode angle $\theta_{k}$ for the parameters of the non-constant-drift RDI simulation
from the main text. As in \cref{fig:linear adep}, different coloured curves show different representative wavenumbers; $k=2\pi/L$ (blue curves), $k=8\pi/L$ (red curves), $k=32\pi/L$ (black curves). Different line styles show different grain-discretisation resolutions -- $N_{d}=256$ (dot dashed), $N_{d}=512$ (dashed), and $N_{d}=1024$ (solid) -- to illustrate the convergence (or lack thereof) to the continuous limit. When the backwards quasi-sound mode
is unstable at $k\,c_{s}t_{s}\lesssim w_{s}^{-1}$ (\cref{eq: k cutoff backwards sound}), this dominates the growth rate for nearly parallel modes and is well converged in $N_{d}$, with a growth rate that is accurately predicted by the analytic 
expression \eqref{eq:backwards sound mode}, shown with the blue dotted line for $k=2\pi/L$. At higher $k$, we see the non-convergence
of the quasi-drift modes (black curves at near-parallel angles) as discussed in \ref{par: quasi drift} and/or the overlapping resonant quasi-sound modes discussed in \ref{par: quasi sound}.  Finally, at $\theta_{k}\approx 85.5$ -- the mode angle resonant with the fastest grains -- we see a converged resonant, forward quasi-sound mode (see inset, which zooms in on this region), with a growth rate that increases with $k$.  This mode is  responsible for the formation of the filamentary structure of large  dust grains seen in \cref{fig:panels.ca}.}\label{fig:linear aindep}
\end{center}
\end{figure}

\subsubsection{Quasi-drift and quasi-sound modes}\label{sub: quasi  modes}
In the shorter wavelength regime, \hs\ identifies two types of (possibly) unstable modes: ``quasi-drift'' modes, so 
named because they are a perturbed drifting dust-density mode with real frequency $\Re(\omega)\approx k_{z} c_{s} w_{s,j}$, and  ``quasi-sound'' modes,
which  are perturbed sound waves with real frequency $\Re(\omega) = \pm k\,c_{s}$. The two modes  behave very differently in the presence of a spectrum of grain velocities, with the quasi-drift modes disappearing in the continuous limit as the density of each grain species tends to zero.
The growth rates of each can be derived by perturbing the eigenvalues of the $\mu=0$ operator,  $\mathbb{T}_{0}=\mathbb{T}|_{\mu=0}$ (see \cref{eq: full matrix}), in the small parameter $\mu\sim \mu_{i}$ with $\mu\mathbb{T}^{(1)}=\mathbb{T}-\mathbb{T}_{0}$. Standard matrix perturbation theory is valid, implying that $\omega=\omega_{0}+\mu\omega^{(1)}$  with $\omega^{(1)}=\xi_{0}^{L}\cdot\mathbb{T}^{(1)}\cdot\xi_{0}^{R}$, where $\xi_{0}^{R}$ ($\xi_{0}^{L}$)
is the right (left) eigenmode of $\mathbb{T}_{0}$ for the (drift or sound) mode that is being perturbed (with $\xi_{0}^{L}\cdot\xi_{0}^{R}=1$).
In the single-grain-species  (or constant drift) instability, the two modes coalesce at the resonant angle, $w_{s}\cos\theta_{k}=1$, where the standard perturbation 
theory results become invalid because of the defective structure of the matrix (see \cref{app: grain dep accel}; \citealp{Squire2018}).
Below, we discuss the quasi-drift then quasi-sound modes in turn,
arguing that the quasi-drift mode disappears in the limit of a continuous grain velocity distribution, while both forward and backward propagating quasi-sound modes remain relevant.
\paragraph{Quasi-drift modes}\label{par: quasi drift}
 The quasi-drift mode is derived by choosing  $\xi_{0}^{R}$ and $\xi_{0}^{L}$ as the drift mode for species $j$, with frequency $\omega_{0}=k_{z} c_{s} w_{s,j}$. Because of the structure of $\mathbb{T}^{(1)}$ and the eigenmodes $\xi_{0}^{R}$ and $\xi_{0}^{L}$, it transpires that 
 the frequency perturbation $\omega^{(1)}$ scales with $\mu_{j}$, \emph{viz.,} the dust-to-gas mass ratio of 
 the \emph{individual} grain species.\footnote{Specifically, the growth rate for the drift mode of species $j$ is
 \begin{equation}
\Im(\omega) = \mu_{j} \frac{1}{t_{s,j}}\frac{\wsj^{2}\cos^{2}\theta_{k}}{\wsj^{2}\cos^{2}\theta_{k}-1}\left( 1- \frac{\zeta_{s,j}}{\tilde{\zeta}_{w,j}}\right),
\end{equation}
which is the same as the single-species result (equation (13) of \hs).
} Importantly, this implies that the growth rate approaches zero as $N_{d}\rightarrow \infty$; instead of a smaller number of fast-growing modes,
 there is a continuum  of modes with an infinitesimally small growth rate. This behaviour is seen for the nearly parallel (small $\theta_{k}$) modes at small scales (black 
 curves) in \cref{fig:linear aindep}:
  the growth rate scales approximately as $\sim\!N_{d}^{-1}$, thus disappearing in the continuous limit. 
 
\paragraph{Quasi-sound modes}\label{par: quasi sound}
The quasi-sound mode is derived by choosing  $\xi_{0}^{R}$ and $\xi_{0}^{L}$ as the eigenmode for a sound wave at $\mu=0$, with frequency $\omega_{0}=k\,c_{s}$. Unlike the quasi-drift modes, the structure of $\xi_{0}^{R}$ and  $\xi_{0}^{L}$ in this case imply that the perturbed 
frequency is the \emph{sum} of the perturbed frequencies from each individual grain species. In other words, if we define $\omega^{(1)}_{j}$ as the frequency 
perturbation to arise if only $\mu_{j}$ is nonzero, with all other $\mu_{i}=0$, then $\omega^{(1)}=\sum_{j}\omega^{(1)}_{j}$. This implies that the 
growth rate (away from resonant regions) scales with $\mu_{0}=\sum_{j}\mu_{j}$ and so can be rapid in the continuum limit. Using
this knowledge, it is straightforward to derive the continuum growth/damping rate of the backwards propagating mode. While the full expression
is complex and unenlightening, it is straightforward to show that the mode is unstable ($\Im(\omega)>0$) for \begin{align}
k\,c_{s}t_{s,j}<&\frac{\wsj^{2}}{(1+\wsj)^{3}[\tilde{\zeta}_{w,j}+\wsj(\zeta_{s,j}-1)]}\tilde{\zeta}_{w,j}^{2}\left(1-\frac{{\zeta}_{s,j}}{\tilde{\zeta}_{w,j}}\right)\nonumber\\ \approx& \frac{1}{\wsj} + \mathcal{L}(\wsj^{-2}) \qquad \text{(Epstein Drag),}\label{eq: k cutoff backwards sound}
\end{align} and a simple expression for 
the unstable regime is found through an expansion in small $k\,c_{s}t_{s}$:
\begin{align}
\Im(\omega)&\approx\frac{1}{2}\sum_{j}\mu_{j}\frac{\wsj^{2}\cos^{2}\theta_{k}}{\tsj(1+\wsj\cos\theta_{k})}\left(1-\frac{\zeta_{s,j}}{\tilde{\zeta}_{w,j}}\right) +\mathcal{O}(k \,c_{s}t_{s}),\nonumber\\
&\approx\frac{1}{2}\int d\ad\,\frac{d\mu}{d\ad}\frac{w_{s}^{2}\cos^{2}\theta_{k}}{t_{s}(1+w_{s}\cos\theta_{k})}\left(1-\frac{\zeta_{s}}{\tilde{\zeta}_{w}}\right).\label{eq:backwards sound mode}
\end{align}
We see that the backwards mode can provide a positive growth rate at wavelengths shorter (up to $k \,c_{s}t_{s,j}\simeq\wsj^{-1}$) than the low-$k$ regime of \cref{sub: low k linear}.\footnote{The expression (10) in \hs\ is the higher-$k$ expression for the same mode.} The expression \eqref{eq:backwards sound mode}
is plotted with the blue-dotted line on top of the numerically computed growth rate for the parameters of the non-constant-drift simulation in \cref{fig:linear aindep}, showing good agreement. By comparison to the nonlinear evolution shown in \cref{fig:panels.ca,fig:time.evolution.ca}, we see  
that this mode is a significant driver of the box-scale instability and causes the formation of the shock wave  in this simulation (see \cref{fig:panels.ca}).

The behaviour of the forward-propagating quasi-sound mode is significantly more complex because 
of the presence of resonances, which overlap with each other due to the distribution of $\wsj$. This leads to 
a complicated dependence of the growth rate on the width of the resonance (which gets narrower with increasing $k\, c_{s}t_{s}$) and
the  distribution of grains. This makes it difficult and subtle to predict analytically, and even to solve
for numerically in the short-wavelength regime,{ features that are shared by the streaming
instability in protoplanetary disks}.  Its basic attributes are best explained through considering 
 the contribution of an individual grain species ($\omega^{(1)}_{j}$), along
with reference to numerically computed growth rate  in \cref{fig:linear aindep}. Like  for the backward quasi-sound mode discussed above,
direct calculation of $\omega^{(1)}_{j}$ yields a complex expression that we do not reproduce in full here. Its  key feature
is that $\Im[\omega^{(1)}_{j}(\theta_{k})]$ is everywhere negative except close the resonant angle $\wsj\cos\theta_{k,j}=1$,
where is has the form
\begin{equation}
\Im(\omega^{(1)})_{j} \propto \frac{1}{1-\wsj \cos\theta_{k}} + \dots.\label{eq: near resonance quasi sound}
\end{equation}
Although the infinity at resonance is spurious (it is regularised by the expansion in $\mu_{j}$ becoming invalid), the general form of \cref{eq: near resonance quasi sound}
shows that quasi-sound modes are very unstable for $\theta_{k}>\theta_{k,j}$ and highly damped for $\theta_{k}<\theta_{k,j}$. The
width of this region of sudden change in $\Im(\omega)$ (i.e., the width of the resonance) decreases with  increasing $k\,c_{s}t_{s}$.
In the presence of a wide spectrum of grains, where the total growth rate is $\sum_{j}\omega^{(1)}_{j}$,
these positive and negative growth rates cancel out between different grains in the regions of $\theta_{k}$ where 
 resonances overlap. This behaviour is seen for $\theta_{k}\lesssim80^{\circ}$ in \cref{fig:linear aindep}, where, despite 
the presence of resonant grains 
the growth rate is unconverged in $N_{d}$ (recall that the non-constant-drift simulation in the main text has grains from $\wsj\approx 1$ to $12$; see \cref{sub: simulation set up}).\footnote{Note that there are also  contributions from the quasi-drift modes at these angles, the growth rate of which also decreases with $N_{d}$ in a similar way.} F{urther discussion of similar issues can be found in \cite{Krapp2019,Krapp2020,Paardekooper2021}.} However,
towards the end of the range of resonant angles, as $\cos\theta_{k}$ approaches $w_{s,{\rm max}}^{-1}$, 
there  no longer exist resonances at higher $\theta_{k}$, which suggests the existence of a converged, resonant growth region around the 
resonance of the largest grains $\cos\theta_{k}w_{s,{\rm max}}=1$. Indeed, this is seen in the inset of \cref{fig:linear aindep}, where we
see fast growth rates around $\theta_{k}\approx 85.5\approx\cos^{-1}(w_{s,{\rm max}}^{-1})$, which are converged to 
the continuum limit (overlapping dot-dashed, dashed, and solid lines). Further, we see that the growth rate increases with $k$, albeit more slowly than
the single-grain case (which scales as $\Im(\omega)\sim k^{1/2}$) because at higher $k$, the resonant width narrows, so there is a smaller 
population of grains that contribute to the converged peak. Thus, we see that a resonant mode 
survives in the presence of a continuous spectrum of grains, although only the largest grains (which stream the fastest) can 
contribute to its growth rate, which is therefore  reduced compared to the constant-drift (or single grain) regime (\cref{fig:linear adep}).


\section{Linear growth in GIZMO}\label{app: linear gizmo comparison}

\begin{figure}
\begin{center}
\includegraphics[width=1.0\columnwidth]{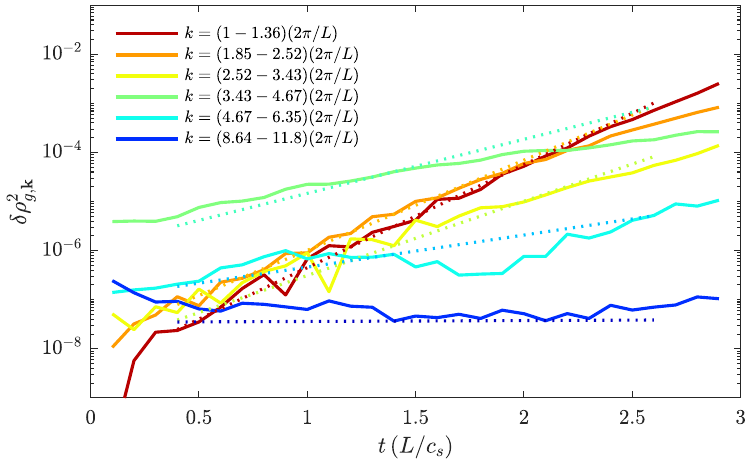}
\caption{{Time evolution of parallel ($\theta_{k}=0$, $\hat{\bm{k}}=\hat{{\bf w}}_{s}$) modes in 
the L1Nd1 simulation (see text). The solid lines show the Fourier energy of gas density density 
perturbations at the labeled $k$ from the box scales (red) to small scales (blue). The dashed lines show
$\exp(2\gamma_{\rm lin}t)$ (arbitrarily rescaled in amplitude), where $\gamma_{\rm lin}$ is computed
from the dispersion relation as described in \cref{app: linear} for the same $k$ and $N_{d}=256$.} }
\label{fig: nl lin comparison large}
\end{center}
\end{figure}

\begin{figure}
\begin{center}
\includegraphics[width=1.0\columnwidth]{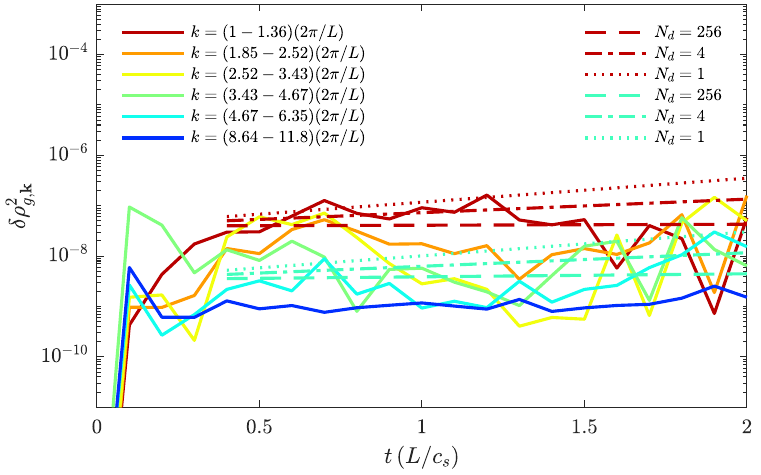}
\caption{{As for \cref{fig: nl lin comparison large}, but for the LsmallNd1, which is identical to L1Nd1 but $12$ times smaller.
The higher $k$ of the box scale suppresses the backward quasi-sound mode (see \cref{fig:linear aindep}),
making the growth rate of all parallel modes effectively zero in the $N_{d}\rightarrow\infty$ limit. The dashed, dot-dashed,
and dotted lines show the linear growth rate of the box-scale modes when computed with different numbers of grain-size bins, as labelled. 
The simulation appears inconsistent with $N_{d}\lesssim 4$, for which there should be a positive growth rate, suggesting
that the continuous grain limit is at least approximately resolved by GIZMO. } }
\label{fig: nl lin comparison small}
\end{center}
\end{figure}

{As discussed above, a number of recent works have explored the difficulty of numerically resolving the linear growth 
of polydisperse dust instabilities in the context of protoplanetary disks (the streaming instability; \citealp{Krapp2019,Paardekooper2021,Zhu2021}). The fundamental problem relates to the discretization of the dust size into $N_{d}$ bins, as there can be very slow (or nonexistent) convergence to
the continuous limit as $N_{d}\rightarrow \infty$. Indeed, the origin of this issue is clear for some of modes discussed in \cref{app: grain indep accel}; for example, the growth rate of  ``quasi-drift modes'' (\cref{sub: quasi  modes}) scales linearly with the relative density of an individual grain size, which approaches zero as $N_{d}\rightarrow \infty$ (see \cref{fig:linear aindep}). As also discussed above (see \cref{fig:linear adep}), this issue is relevant only for the non-constant-drift regime, 
and in the constant-drift regime, linear results are well converged even at relatively small $N_{d}$. }

{In this appendix, we assess the relevance of such effects to our nonlinear simulations, by computing  linear 
growth rates in their early phases. We focus on only the non-constant-drift case and the particular concern 
of resolving spuriously large growth rates due to insufficient resolution in dust size space.  The method 
also acts as a stringent test of the ability of GIZMO to resolve the linear physics of the RDI more generally. 
Note, however, that because this method uses the nonlinear results to probe the linear physics, modes can
in principle interact; for example, a mode that is linearly neutral or damped may grow due to the nonlinear interactions 
even at small amplitudes. In addition, meshless methods such as that used by GIZMO are inherently relatively
noisy for low amplitude fluctuations, which implies that the linear growth phase starts from relatively 
larger amplitudes (compared to a fixed-mesh numerical method) and makes resolving the linear phase more difficult. }

{Our method involves using a Fourier decomposition of the gas density in the simulation to compute 
the time evolution of individual Fourier modes. We then bin these in $k=|\bm{k}|$ and $\theta_{k} = \cos^{-1}(k_{z}/k)$
to compare to the predicted rates at a given scale and mode angle. As can be seen from \cref{fig:linear aindep}, the 
non-convergent linear behavior manifests itself only at rather small scales, approximately 10 times
smaller than the box scale and smaller. Thus, 
in order to correctly assess the effect and ensure that it is not unduly influenced by numerical dissipation between
gas elements, we have run two extra simulations. The first, which we label L1Nd1, is identical to the 
non-constant-drift case in the main text, with box size $L_{0}=1$ in code units, but with $128^{3}$ gas resolution elements and the same 
number of dust elements. The second, labeled LsmallNd1, has a box size $L_{\rm small}$ that is $12$ times
smaller, $L_{\rm small}=L_{0}/12$, so as to more directly resolve the small-scale dynamics (this also makes it significantly more computationally expensive). 
We stress that because these simulations use the same number of dust and gas elements, they should 
be a {more} stringent test of the features we wish to study (non-convergence in $N_{d}$) compared to 
the simulations of the main text, which used $4\times256^{3}$ dust elements and $256^{3}$ gas elements (the purpose
of the L1Nd1 simulation is to more appropriately compare to the LsmallNd1 simulation). As for the
simulations in the main text, the dust is discretized in size by sampling from a continuous distribution. }

\begin{figure}
\begin{center}
\includegraphics[width=1.0\columnwidth]{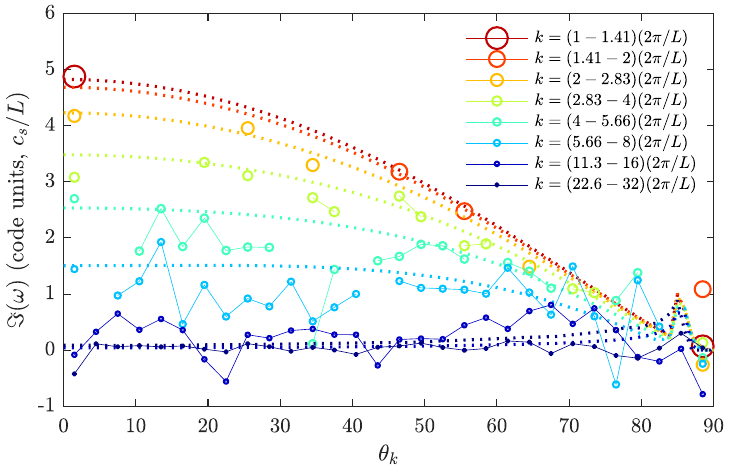}
\caption{{Points and solid lines show the growth rates as a function of $k=|\bm{k}|$ and $\theta_{k}$ measured from the high-resolution non-constant-drift simulation 
discussed in the main text. We measure each growth rate by fitting each Fourier mode of the gas density to an exponential in time 
for $t\leq3$. At larger scales, there are many fewer Fourier mode angles available because of the discretization of $\bm{k}$ in the box, which 
explains the sparse points for larger scale modes (red colors). The dashed lines show the linearly computed growth rate as a function 
of $\theta_{k}$ (as also shown in \cref{fig:linear aindep}) for each of the bins in $k$.  }}
\label{fig: nl lin comparison k theta}
\end{center}
\end{figure}

{The results of these two simulations are shown in \cref{fig: nl lin comparison large,fig: nl lin comparison small}.
We illustrate just the parallel modes here -- i.e., $\theta_{k}=0$ or $\hat{\bm{k}}=\hat{{\bf w}}_{s}$ -- which
are the ``backwards quasi-sound'' and/or ``quasi-drift'' modes at these parameters. Comparing the
numerical (solid) and linear (dotted) results in \cref{fig: nl lin comparison large}, we see a very good match, showing that 
GIZMO resolves the linear growth in the expected way, although the nonlinear results are not perfect exponentials due to 
the noise inherent in the meshless method of GIZMO (fluctuation amplitudes remain  small at this time, so the noise makes a relatively larger contribution). Similar results from  LsmallNd1 are shown in \cref{fig: nl lin comparison small}. The converged linear growth rates
of these parallel modes ($\theta_{k}=0$) are effectively zero  for all modes, $k>12\times 2\pi/L$, that fit in the box. In order to assess convergence 
we plot the linear prediction given different discretizations in dust size space, as discussed in \cref{sub: quasi  modes}. 
Although the linearly predicted growth rates are very slow even for $N_{d}=1$, and so are hard to tell apart in the somewhat noisy numerical data, the simulation results show no indication of linear growth and thus appear inconsistent
with linear predictions for $N_{d}\leq 4$. Given that this simulation is discretized with one dust particle per gas particle, this is
reasonable evidence that the key physics of the continuous grain spectrum is being resolved by our method. }


{It has proven somewhat more difficult to assess the convergence of the growth rate for the overlapping quasi-resonant  ``forward quasi-sound modes'' 
at these parameters, although we clearly see nearly perpendicular modes growing strongly 
in large dust grains in  both the high-resolution main-text simulation  (see \cref{fig:time.evolution.ca}) and the LsmallNd1 simulation, and these broadly 
match the predicted growth rates.   We speculate that the difficulty of measuring
the detailed mode structure relates to its relative narrowness in $\theta_{k}$ and the influence of the remeshing
noise. Given the strong sensitivity of even the linear results to $k$ and $N_{d}$, a detailed study of the linear growth 
of these modes would require other simulations with careful convergence checks in $N_{d}$ and $k$, and is beyond the
scope of this work (see \citealt{Krapp2019,Paardekooper2021,Zhu2021}). In \cref{fig: nl lin comparison k theta}, we show the growth rate as a function of $\theta_{k}$ and $k$, measured 
from the non-constant-drift simulation of the main text for $t\leq3$. The agreement with linear predictions is good for quasi-parallel modes, and
reasonable for quasi-perpendicular modes, most  of which grow around the linearly predicted rate (note
that for a given $k$, modes exist only for certain $\theta_{k}$ due to the grid, which makes it hard to resolve the fine structure).  Given the uncertainties
in the measurement method -- different modes can interact, and the simulation is somewhat noisy in this  low-amplitude initial phase -- 
the overall agreement is reasonable, and  justifies the ability of GIZMO to resolve the mode structure of the polydisperse acoustic RDI.}

\end{document}